\documentclass[sigconf]{acmart}

\AtBeginDocument{%
  \providecommand\BibTeX{{%
    \normalfont B\kern-0.5em{\scshape i\kern-0.25em b}\kern-0.8em\TeX}}}

\setcopyright{acmcopyright}
\copyrightyear{2022}
\acmYear{2022}
\acmDOI{XX.XXXX/XXXXXXX.XXXXXXX}

\acmConference[XX]{XX}{XX}{XX}
\acmBooktitle{XX,X,X}
\acmPrice{15.00}
\acmISBN{XXX-X-XXXX-XXXX-X/22/07}

\usepackage{algorithm}
\usepackage{algorithmic}
\usepackage{multirow}
\usepackage{subfigure}
\usepackage{enumitem}

\begin{document}

\title{UFNRec: Utilizing False Negative Samples for Sequential Recommendation }



\author{Xiaoyang Liu\textsuperscript{\normalfont 1}$^{\ast}$, Chong Liu\textsuperscript{\normalfont 2}}\authornote{Equal Contribution.}\author{Pinzheng Wang\textsuperscript{\normalfont 3}, Rongqin Zheng\textsuperscript{\normalfont 2}, Lixin Zhang\textsuperscript{\normalfont 2}, Leyu Lin\textsuperscript{\normalfont 2}, Zhijun Chen\textsuperscript{\normalfont 1}, Liangliang Fu\textsuperscript{\normalfont 1}}
\email{liuxiaoyang@oppo.com}
\email{nickcliu@tencent.com}
\affiliation{%
  \institution{\textsuperscript{\normalfont 1}OPPO, \textsuperscript{\normalfont 2}Tencent, \textsuperscript{\normalfont 3}Soochow University}
  \country{}
  }

\renewcommand{\shortauthors}{Xiaoyang Liu and Chong Liu, et al.}

\begin{abstract}
Sequential recommendation models are primarily optimized to distinguish positive samples from negative ones during training in which negative sampling serves as an essential component in learning the evolving user preferences through historical records. 
Except for randomly sampling negative samples from a uniformly distributed subset, many delicate methods have been proposed to mine negative samples with high quality.
However, due to the inherent randomness of negative sampling, false negative samples are inevitably collected in model training.
Current strategies mainly focus on removing such false negative samples, which leads to overlooking potential user interests, lack of recommendation diversity, less model robustness, and suffering from exposure bias. To this end, we propose a novel method that can \textbf{U}tilize \textbf{F}alse \textbf{N}egative samples for sequential \textbf{Rec}ommendation (UFNRec) to improve model performance.
We first devise a simple strategy to extract false negative samples and then transfer these samples to positive samples in the following training process.
Furthermore, we construct a teacher model to provide soft labels for false negative samples and design a consistency loss to regularize the predictions of these samples from the student model and the teacher model.
To the best of our knowledge, this is the first work to utilize false negative samples instead of simply removing them for the sequential recommendation.
Experiments on three benchmark public datasets are conducted using three widely applied SOTA models. 
The experiment results demonstrate that our proposed UFNRec can effectively draw information from false negative samples and further improve the performance of SOTA models.
The code is available at \url{https://github.com/UFNRec-code/UFNRec}.
\end{abstract}




\maketitle

\section{Introduction}
Recommendation systems play an essential role in online platforms (e.g., Amazon~\cite{amazon}, Facebook~\cite{ebr} and Google~\cite{youtubeDNN}) to learn evolving user preferences and fulfill user requirements. 
Generally, the core of the sequential recommendation task is to optimize neural models that can distinguish positive samples from negative ones based on users’ historical iterations.
Many methods have been proposed to capture user interests and recommend items accurately, including RNN-Based methods~\cite{GRU4Rec,GRU4Rec+}, GNNs structures~\cite{SR-GNN,FGNN,GC-SAN}, various CNNs frameworks~\cite{Caser,3D_CNN}, and model variants~\cite{BERT4Rec,TiSASRec,sasrec} based on the effective multi-head self-attention~\cite{atten}.

Though achieving promising results, these methods commonly apply items with implicit historical feedback (e.g., clicks) of users as positive samples and regard items randomly selected from a specific set, e.g., the training dataset or a mini-batch samples, as negative samples. Thus, these methods significantly rely on the quality of negative samples and roughly assume that items without historical interactions are negative samples.
However, due to the inherent randomness of negative sampling, false negative samples naturally exist in negative samples, which means a selected negative item can be clicked by the user in the future.
For example, the items exposed to a user are entirely determined by the recommendation system, leading to exposure bias. Thus, if an item is never recommended to the user, he/she will never have the opportunity to interact with this item. In other words, this item can be interacted with by the user if exposed.
Such false negative samples can affect the recommendation quality and model robustness, especially for some data-sensitive models. 
Furthermore, since the false negative samples are potential positive samples, overlooking these false negative samples will lead the model to learn the narrow interests of users, which influences the recommendation diversity. Thus, a recommendation system should be able to distinguish false negative samples from all negative samples and further utilize them.

However, it is non-trivial to devise an effective method to utilize false negative samples in the following two noteworthy aspects:
1) Due to the randomness of negative samples and the lack of reliable supervised signals for false negative samples, it is challenging to design a valid criterion to distinguish false negative samples from all randomly selected negative ones.
Significantly, such a criterion should be instance-level and context-aware. 
For instance, if a negative sample fits the interests of the corresponding user or is highly related to the positive item, this sample can be a false negative sample.
2) Because of the inherent uncertainty of such false negative samples, we should carefully design an approach to utilize these samples. On the one hand, simply removing these samples can avoid introducing noise but roughly ignore the significant benefits of such samples. On the other hand, a loose criterion will bring additional noise and lead to inferior model performance. Thus, the strategy to utilize false negative samples should be thorough enough.

Many studies have paid attention to the influence of false negative samples~\cite{zhao2018interpreting,hernandez2014probabilistic}. Early studies~\cite{liang2016modeling} rely on extra information such as item content to identify possible false negative samples and train them with low weights. 
Recently, SRNS~\cite{ding2020simplify} leverages a variance-based sampling strategy to remove false negative samples.
Besides, studies in Knowledge Graph (KG) also find that introducing false negative samples will lead to inferior model performance. With the observation that the scores of false negative samples in KG can be very high, NSCaching~\cite{zhang2019nscaching} directly removes the sample with the largest score in each negative sample set. 
Though promising, these methods only focus on reducing the risk of false negative samples without explicitly utilizing them, which overlooks the importance of false negative samples in enhancing user experience.

To this end, in this paper, we propose an effective strategy to utilize false negative samples for sequential recommendation. 
Firstly, we leverage the item scores to filter out false negative samples from negative samples without involving extra data. Concretely, we record negative samples with larger scores than positive ones and consider negative samples recorded over certain times (i,e,.reaching a given threshold) as false negative samples. 
Unlike SRNS~\cite{ding2020simplify} that simply removes false negative samples, we utilize these samples to further train the model. 
Specially, we reverse the labels of these samples and treat them as positive samples during the following training process.
In this way, we distinguish the potential positive items from the original randomly sampled negative items, which can widen user interests and improve the recommendation diversity. 
Furthermore, to enhance model robustness, we adopt a consistency training loss on a teacher-student framework to regularize the output distribution of newly exploited false negative samples. 
Experiments on three backbone models and three real-world datasets prove that our proposed UFNRec can improve model performance for sequential recommendation over different state-of-the-art baselines. 
This work mainly includes the following contributions:
\begin{itemize} 
\item We propose an effective strategy for sequential recommendation systems to identify false negative samples and transfer them to positive samples to train the model further.
To our best knowledge, this is the first work to utilize false negative samples to improve recommendation performance.
\item We introduce a consistency training method based on a teacher-student framework to regularize the predictions of false negative samples.
\item Extensive experiments conducted with three SOTA models on three widely used public datasets indicate the superiority of our proposed UFNRec over SOTA models.
\end{itemize}

\section{Related Work}
\subsection{Sequential Recommendation(SR)}
Early works widely apply Markov Chain (MC) to model sequential interactions, including both the first-order MC~\cite{FPMC} and the high-order MC~\cite{TransRec,Fossil}. With the prosperity of deep learning approaches, CNN-based~\cite{Caser,3D_CNN,RCNN}, RNN-based~\cite{GRU4Rec,NARM,HRNN,GRU4Rec+,jing2017neural,liu2016context,LSTM},
attention-based~\cite{sasrec,BERT4Rec,TiSASRec,SHAN,MARank,ATRank}, and GNN-based~\cite{SR-GNN,FGNN,GC-SAN,GCN,APP-GE,wang2020make} methods are proposed to capture user interests and enhance model performance. 
For example, GRU4Rec~\cite{GRU4Rec} utilized an RNN-based method for the SR task, and then many subsequent works have been proposed based on GRU4Rec~\cite{GRU4Rec+,HRNN,Air,NARM,SDM,lossRNN,UserBased,Latent_Cross}. 
However, RNN-based methods are usually more sensitive to the problem of data sparsity and thus perform worse than  CNN-based and attention-based methods. 
As for CNN-based methods, Caser~\cite{Caser} applied CNN to extract users' short-term preferences, and \citet{3D_CNN} used CNN to combine content features and session clicks. Then, NextItNet~\cite{NextItNet} designed dilated convolutions to comprehend long-range sequences. 
Also, many researchers introduced attention mechanisms~\cite{atten} to improve recommendation performance and achieved great success~\cite{sasrec,BERT4Rec,TiSASRec,E-BART4Rec,fan2021lighter,liu2021augmenting} in the SR task.
For example, SASRec~\citet{sasrec} introduced a self-attention mechanism to identify relevant items from users' interaction history and recommend the next item.
BERT4Rec~\cite{BERT4Rec} applied the bidirectional self-attention to model user interaction sequences.
Moreover, combining attention mechanisms with GNNs~\cite{SR-GNN,FGNN,GC-SAN} proves to be another alternative way for the SR task. Some methods like transfer learning and data augmentation also show their efficacy in improving the performance of SR~\cite{MANN,KV-MN,disentangled1,GLaS,SSL,MIRec,2017hu}.

Recently, contrastive learning is another effective way to solve the SR task. Specially, methods that combine contrastive learning and attention mechanisms~\cite{CL4SRec,CLRec,CauseRec} have achieved significant success. Concretely, CL4SRec~\cite{CL4SRec} obtains self-supervision signals based on data augmentation methods and applies a contrastive loss to restrict the model outputs. 
Besides, StackRec~\cite{StackRec} introduces a highly deep but easy-to-train model by stacking a pre-trained shallow model and fine-tuning it. ICAI-SR~\cite{ICAI-SR} introduces how to handle the complex relations between items and categorical attributes, and SINE~\cite{SINE} focuses on extracting multiple interests of users for the SR task. Also, CT4Rec \cite{CT4Rec} proposes an effective consistency training method by only adding two extra training objectives, and this method obtains outstanding achievement without any structural modifications or data augmentation strategies.

\subsection{Negative Sampling for Sequential Recommendation}
\begin{figure*}[t]
    \centering
    \includegraphics[scale=0.33]{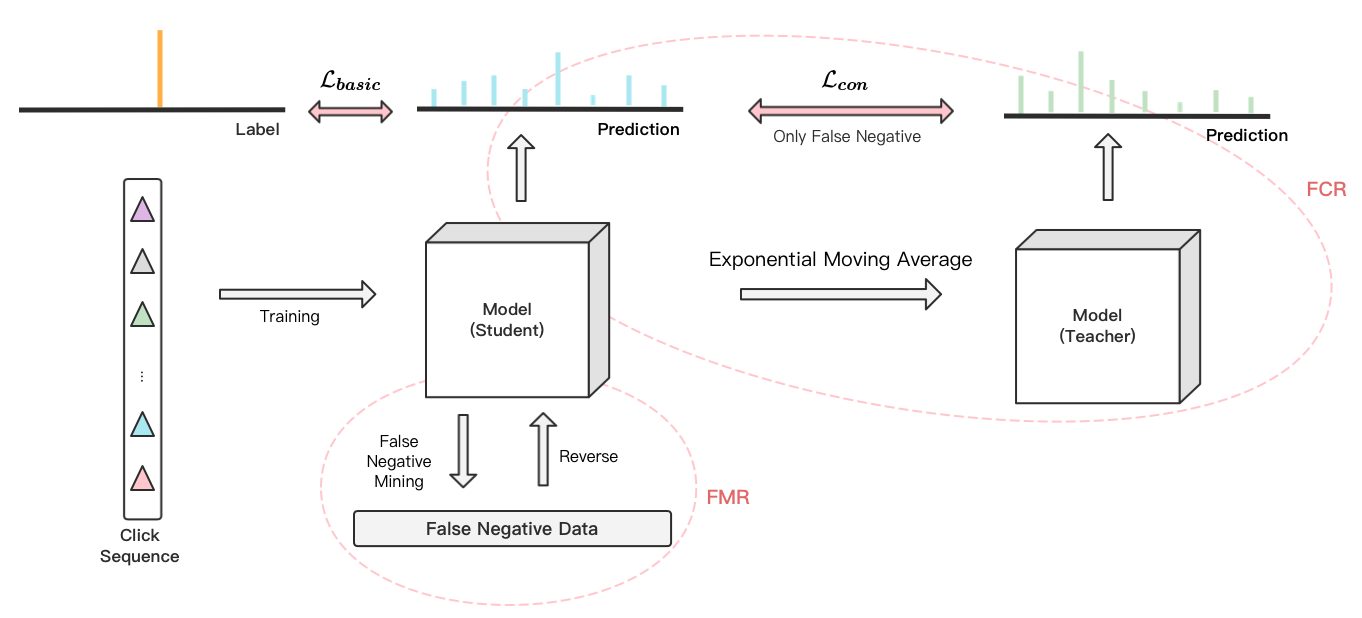}
    \caption{Model structure of UFNRec. It filters out false negative samples and reverses their labels to train the model. In addition, a teacher model is generated from current student models by EMA, and a consistency loss is introduced to regularize outputs from the teacher-student framework.}  
    \label{Fig:main}
    \vspace{-0.3cm}
    \end{figure*} 
Many sampling-based methods ~\cite{bengio2008adaptive,pmlr-vR4-bengio03a,blanc2018adaptive,mikolov2013distributed} have been used to improve the model performance for SR. Among them, some studies~\cite{hard-1,hard-2,hard-3,hard-4} pay attention to the selection of hard negative samples. 
Concretely, ANCE~\cite{xiong2020approximate} utilized an asynchronously updated ANN index to select hard negative samples globally. Besides, the efficiency of negative sampling~\cite{youtubeDNN,chen2017sampling} has recently become a research hotspot in recommendation tasks. 
Specifically, MNS~\cite{yang2020mixed} mixed uniformly and in-batch negative samples to alleviate the selection bias. CBNS~\cite{wang2021cross} focused on cross batch negative sampling strategy instead of only sampling negative data in an in-batch way. \cite{yi2019sampling} corrected the sampling bias by estimating item frequency through streaming data. 

However, these methods mainly concentrate on effectively mining negative samples and ignore the existence of false negative samples~\cite{hernandez2014probabilistic}. Few studies notice the influence of false negative samples for the sequential recommendation.
\cite{liang2016modeling} identified possible false negative samples by additional information and then reduced the training weights of these samples. 
\cite{zhao2018interpreting} interpreted the complicated and varied reasons for users' inaction and emphasized that inaction information brings benefits to recommendation systems. 
In the domain of knowledge graph, NSCaching~\cite{zhang2019nscaching} found that the scores of false negative samples can be very high and thus removed the sample with the largest score in each negative sample set. 
For recommendation tasks, SRNS~\cite{ding2020simplify} recently proposed a variance-based strategy to filter out false negative samples and removed these samples during training. Though achieving strong performance, these methods only focus on how to distinguish false negative samples from all negative samples and then reduce the risk of introducing these samples. Different from their studies, we address the utilization of false negative samples, which can improve model robustness and recommendation diversity.

\section{The UFNRec Model} \label{sec:method}
Figure~\ref{Fig:main} shows the overall architecture of our model.
Before fully explaining UFNRec, we introduce some basic notations to describe the sequential recommendation task. Let \(\mathcal{V}=(v_1,v_2,...,v_{|\mathcal{V}|})\) and \(\mathcal{U}=(u_1,u_2,...,u_{|\mathcal{U}|})\) respectively denote a group of items and users. For a given user \(u\in\mathcal{U}\), the historical interaction sequence is denoted as \(s_u=(v_1^{(u)},v_2^{(u)},...,v_t^{(u)},...,v_{|s_u|}^{(u)})\), where \(v_t^{(u)}\in\mathcal{V}\) is the item interacted with user $u$ at time step $t$ and \(|s_u|\) denotes the length of \(s_u\). 
The goal of sequential recommendation is to predict the probability of all alternative items based on the historical sequence \(s_u\) and then recommend the most likely item that the user $u$ will interact with at time step \(|s_u|+1\). This prediction task can be formulated as below:
\begin{equation}
    \begin{split}
        {P}(v_{|s_u|+1}^{(u)}=v|s_u )
      \label{func:bs}
    \end{split}
\end{equation}
\subsection{Backbone Model}
Since our proposed UFNRec method does not depend on structural information, we choose three widely used models (i.e., SASRec~\cite{sasrec}, BERT4Rec~\cite{BERT4Rec} and SSE-PT~\cite{wu2020sse}) as backbones. Generally, user representation \(\boldsymbol{s_u}=f(s_u)\) can be obtained at each time step $t$, where \(f(\cdot)\) indicates the model encoder.

To learn the correlation between users and items in the sequential recommendation, a similarity function, e.g., inner product, is used to measure the distance between item representation and user representation. 
Then, as shown in Equation~\ref{func:basic_pro}, a binary cross entropy loss function \(l_{u,t}(\boldsymbol{v_{t+1}})=l(\boldsymbol{v_{t+1}}|\boldsymbol{s_{u,t}})\) can be inferred for the user representation \(\boldsymbol{s_{u,t}}\) of user $u$ at time step $t$. 
Here, $\sigma(\boldsymbol{s_{u,t}v_{t+1}})$ is the prediction result and $y$ is the label of $v_{t+1}$, with $y=1$ for positive samples and $y=0$ for negative ones.
\begin{equation}
    \begin{split}
    l_{u,t}(\boldsymbol{v_{t+1}})= -y log(\sigma(\boldsymbol{s_{u,t}v_{t+1}})) 
    - (1-y)log(1-\sigma(\boldsymbol{s_{u,t} v_{t+1}}))  \label{func:basic_pro}
    \end{split}
\end{equation}
\begin{equation}
    \begin{split}
    \sigma(x)=1/(1+e^{-x})  \label{func:$sigmoid$}
    \end{split}
\end{equation}
Commonly, for each positive sample, $n$ negative items are randomly selected from the dataset to obtain a set $\mathcal{N}$ with $|\mathcal{N}|=n$. Then, a basic loss is utilized to measure the prediction accuracy for the positive item \(\boldsymbol{ v_{t+1}^+}\) and $n$ negative items \(\boldsymbol{ v_{t+1}^-}\) from the set $\mathcal{N}$:
\begin{equation}
    \begin{split}
    \mathcal{L}_{basic}(\boldsymbol{s_{u,t}};\omega)= l_{u,t}(\boldsymbol{ v_{t+1}^+};\omega) + \sum\limits_{\boldsymbol{ v_{t+1}^-}\in{\mathcal{N}}}l_{u,t}(\boldsymbol{ v_{t+1}^-};\omega)
    \label{func:basic_loss}
    \end{split}
\end{equation}

\subsection{False Negative Mining and Reversing (FMR)}
\label{sec:method_FMR}
As mentioned above, the basic loss of SR models relies on both positive and negative samples. Due to the randomness of these $n$ negative samples from \({\mathcal{V}}\), the quality of set $\mathcal{N}$ is doubtful. Specifically, some negative samples can be false negative samples. For example, it is possible that a randomly selected item is a potential positive sample but regarded as a negative sample during the training process. 
Thus, without discriminating false negative samples, the model will train with both true negative samples and false negative samples, making the model less robust. 
To address this problem, we design an efficient strategy to filter out false negative samples. Moreover, we utilize false negative samples to further train the model by reversing the labels of these samples. 
\begin{figure}[t]
    \centering
    \includegraphics[scale=0.315]{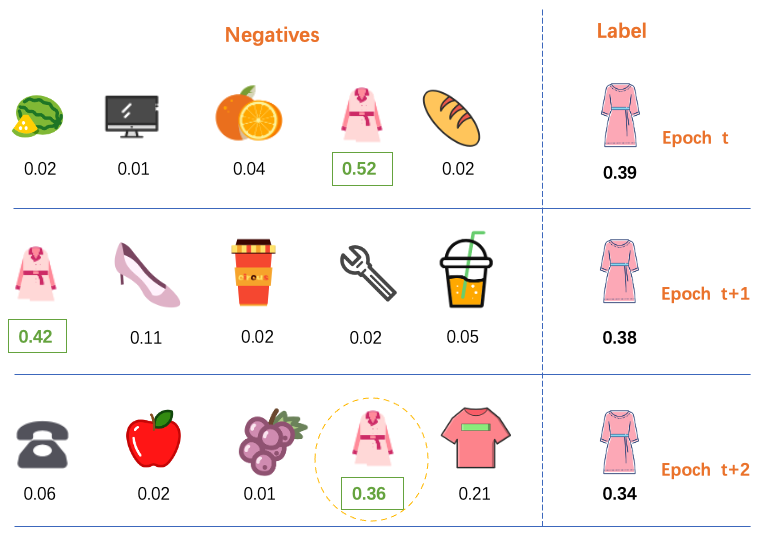}
    \caption{An illustrative example of how to mine a false negative sample from randomly selected negative samples.}  
    \label{Fig:false}
    \end{figure} 
    
\noindent\textbf{False Negative Mining.}
Motivated by the observation that false negative samples always have larger scores than true negative samples~\cite{zhang2019nscaching}, we identify false negative samples according to their prediction results.
Before applying our strategy, we firstly warm up the model by utilizing the randomly sampling strategy until the basic loss converges. After that, we start to mine false negative samples. 
Concretely, during each epoch, we record negative samples (i.e., set $\mathcal{N}_{rec}$) that have larger prediction results than positive samples (i.e., $\sigma(\boldsymbol{s_{u,t}v_{t+1}})$). Moreover, we retrain them with other randomly sampled negative items (i.e., set $\mathcal{N}_{ran}$) in the next epoch following Equation~\ref{func:basic_loss_uni} with ${v_{t+1}^-\in{\mathcal{N}_{ran}}\cup\mathcal{N}_{rec}}$ and $|\mathcal{N}_{ran}|=n-|\mathcal{N}_{rec}|$. 
\begin{equation}
    \begin{split}
    \mathcal{L}_{basic}(\boldsymbol{s_{u,t}};\omega)= l_{u,t}(\boldsymbol{ v_{t+1}^+};\omega) + \sum\limits_{\boldsymbol{ v_{t+1}^-}\in{\mathcal{N}_{ran}}\cup\mathcal{N}_{rec}}l_{u,t}(\boldsymbol{ v_{t+1}^-};\omega)
    \label{func:basic_loss_uni}
    \end{split}
\end{equation}
Then, we distinguish false negative samples from set $\mathcal{N}_{rec}$ which consists of potential false negative samples.
Concretely, when negative samples are recorded over $m$ epochs (a predefined hyper-parameter), we suppose these negative samples to be false negative samples (i.e., set $\mathcal{N}_{false}$ ). 
Here, the parameter $m$ approximately indicates the benchmark to identify false negative samples. A larger $m$ means a stricter rule for false negative samples and vice versa.
Thus, if a negative sample obtains higher scores than the corresponding positive item over $m$ times, we can conclude this sample is a potential positive sample.
Figure~\ref{Fig:false} shows how to mine false negative samples through their scores with $m=3$.

\noindent\textbf{False Negative Reversing.}
Recent researches~\cite{ding2020simplify,zhang2019nscaching} commonly treat false negative samples as noise and focus on reducing the risk of involving these samples. However, we suppose these false negative samples can provide undiscovered user interests. Thus, we believe that utilizing these false negative samples can achieve better model performance than removing them.
Specially, we reverse the labels of samples in $\mathcal{N}_{false}$ and train the model following Equation~\ref{func:basic_loss} with $\boldsymbol{v_{t+1}^+}\in{\mathcal{N}_{false}}$.
Here, we utilize false negative samples for sequential recommendation in a simple but effective way.
\subsection{False Negative Consistency Regularization (FCR)}
\label{sec:method_FCR}
\noindent\textbf{Exponential Moving Average ($\boldsymbol{EMA}$).}
Besides simply training false negative samples as positive ones, we derive soft labels for these samples to improve label quality.
Considering that an ensemble of historical models is more stable and accurate than a single final model~\cite{2017Mean}, we apply a teacher-student framework to construct soft labels.
Models updated with gradient descent over each batch are regarded as student models, and an ensemble of student models generates a teacher model.
Instead of directly averaging weights, the teacher model applies the Exponential Moving Averaging ($\boldsymbol{EMA}$) weights of student models to emphasize the influence of newly updated student models.
Concretely, the parameters $\omega'_{t}$ of the teacher model at time step $t$ are updated from the corresponding student model parameters $\omega_{t}$ by $\boldsymbol{EMA}$:
\begin{equation}
    \begin{split}
    \omega'_{t}= d\omega'_{t-1} + (1-d) \omega_{t}
    \label{func:tea_update}
    \end{split}
\end{equation}
where $d \in [0,1]$ is a smoothing rate.
In this way, we effectively introduce a teacher model from student models without any changes to the original model structures. 
Also, the teacher model can aggregate representations from current student models and provide stable soft labels for false negative samples. 

\noindent\textbf{Consistency Regularization Loss.}
Moreover, for false negative samples, we propose a consistency training method to regularize the outputs of the teacher model and the student model.
For $\boldsymbol{v_{t+1}^+}\in{\mathcal{N}_{false}}$, we can obtain predictions from the teacher model and regard these predictions as soft labels $\hat{y}$ for corresponding outputs from the student model. Then, a consistency loss is devised to regularize these outputs following Equation~\ref{func:con_pro} and Equation~\ref{func:con_loss}.
\begin{equation}
    \begin{split}
    \widehat{l}_{u,t}(\boldsymbol{v_{t+1}};\omega)= -\hat{y} log(\sigma(\boldsymbol{s_{u,t}v_{t+1}})) 
    - (1-\hat{y})log(1-\sigma(\boldsymbol{s_{u,t} v_{t+1}}))  \label{func:con_pro}
    \end{split}
\end{equation}
\begin{equation}
    \begin{split}
    \mathcal{L}_{con}(\boldsymbol{s_{u,t}};\omega)= \sum\limits_{\boldsymbol{ v_{t+1}^+}\in{\mathcal{N}_{false}}}\widehat{l}_{u,t}(\boldsymbol{ v_{t+1}^+};\omega)
    \label{func:con_loss}
    \end{split}
\end{equation}


\subsection{Final Objective.} 
Finally, we train the above consistency objective together with the backbone model's basic loss.
The final objective is defined as:
\begin{equation}
    \mathcal{L}_{final}= \mathcal{L}_{basic} + \alpha \mathcal{L}_{con} \label{func:final_loss}
\end{equation}
where \(\alpha\) is the coefficient weight to control the effect of \(\mathcal{L}_{con}\). And the student model parameters are updated with the $\mathcal{L}_{final}$:
\begin{equation}
    \begin{split}
    \omega=\omega - \gamma\frac{\partial\mathcal{L}_{final}}{\partial\omega}
    \label{func:stu_update}
    \end{split}
\end{equation}
,where $\gamma$ is the learning rate.

\subsection{Training Algorithm}
The whole training process of UFNRec is shown in Algorithm \ref{alg:alg}. 
As shown in line 2-7, we only train the $\mathcal{L}_{basic}$ while the basic loss descends rapidly. 
Then, we start to generate teacher models following Equation~\ref{func:tea_update} (line 9).  
Here, negative samples are firstly collected from $\mathcal{N}_{rec}$ and then randomly sampled from the whole dataset (line 11).
We filter out negative samples that have larger scores than related positive samples and update $\mathcal{N}_{rec}$ (line 12).
After that, we regard negative samples recorded over $m$ times as false negative samples and update $\mathcal{N}_{false}$ (line 13).
Line 14-15 calculate \(\mathcal{L}_{con}\) only for $\boldsymbol{v_{t+1}^+}\in{\mathcal{N}_{false}}$ and calculate \(\mathcal{L}_{basic}\) for all positive samples including $\boldsymbol{v_{t+1}^+}\in{\mathcal{N}_{false}}$.
Line 16-17 compute \(\mathcal{L}_{final}\) based on Equation~\ref{func:final_loss} and update the model parameters. The whole training process will continue until convergence.

\begin{algorithm}
	\renewcommand{\algorithmicrequire}{\textbf{Input:}}
	\renewcommand{\algorithmicensure}{\textbf{Output:}}
	\caption{UFNRec Training Algorithm }
	\label{alg:alg}
	\begin{algorithmic}[1]
		\REQUIRE Training data  $\mathcal{D}=\left\{s_{u_i,t}\right\}^N_{i=1} $
		\ENSURE  model parameters $\omega$
		\STATE Initialization model with parameters $\omega$
		\STATE Initialization empty set $\mathcal{N}_{rec}$
		
		\WHILE{$\mathcal{L}_{basic}$ descending rapidly}
		\STATE $s_{u_i,t} \sim \mathcal{D} $
		\STATE randomly sample $\boldsymbol{ v_{t+1}}^-\in\mathcal{V} $
		\STATE $  g  \leftarrow \bigtriangledown_\omega \mathcal{L}_{basic}$
		\STATE $\omega \leftarrow GradientUpdate(\omega,g) $ 
		\ENDWHILE 
	    \WHILE{not converged}
		\STATE $s_{u_i,t} \sim \mathcal{D} $
		\STATE sample $\boldsymbol{v_{t+1}^-}\in{\mathcal{N}_{ran}}\cup\mathcal{N}_{rec}$
		\STATE update $\mathcal{N}_{rec}$ based on prediction results $\sigma(\boldsymbol{s_{u,t}v_{t+1}})$
		\STATE update $\mathcal{N}_{false}$ based on parameter $m$
		\STATE update $\mathcal{L}_{basic}$ based on Equation~\ref{func:basic_loss} including $\boldsymbol{v_{t+1}^+}\in{\mathcal{N}_{false}}$
		\STATE update teacher model parameters $\omega'$ with Equation~\ref{func:tea_update} 
		\STATE update $\mathcal{L}_{con}$ based on Equation~\ref{func:con_loss} for $\boldsymbol{v_{t+1}^+}\in{\mathcal{N}_{false}}$
		\STATE $  g \leftarrow \bigtriangledown_\omega \mathcal{L}_{final}$
		\STATE $\omega \leftarrow GradientUpdate(\omega,g) $ 
		\ENDWHILE 
	\end{algorithmic}  
\end{algorithm}

\section{Experiments}
To deeply understand UNFRec, we conduct extensive experiments to answer the following questions:
\begin{itemize}
\item[RQ1] Does UFNRec improve the performance of state-of-art methods for the sequential recommendation? 
\item[RQ2] Do different components of UFNRec benefit the model performance, i.e., FMR and FCR? 
\item[RQ3] How do different false negative mining methods perform?
\item[RQ4] What is the effect of removing false negative samples compared with utilizing them?
\item[RQ5] What is the influence of essential hyper-parameters on UFNRec performance? 
\item[RQ6] How does UFNRec impact the model performance during the training process? 
\item[RQ7] Is there any illustrative example to explain the effect of UFNRec compared with other state-of-art methods? 
\end{itemize}
\subsection{Datasets}
Extensive experiments have been conducted on three widely used public benchmark datasets. We present their detailed statistics in Table~\ref{tb1}. 

\noindent\textbf{Amazon.}
Datasets introduced by \cite{amazon_datasets} product reviews collected from \textit{Amazon.com}. Top-level product categories are used to separate these datasets, and we utilize the \textbf{\textit{Beauty}} and \textbf{\textit{Sports}} categories to evaluate model performance.

\noindent\textbf{Yelp.}
is a well-known dataset for business recommendations, which is culled from the Yelp platform\footnote{https://www.yelp.com/dataset}. Following \citet{s3}, we only leverage data after January 1st, 2019. Also, business categories are regarded as attributes.

\begin{table}[t]
    \caption{Dataset statistics of three public benchmarks, where \textit{avg.} refers to the average actions per user.}
    \centering
    \renewcommand\arraystretch{1.06}
    \begin{tabular}{cccccc}
    \toprule
         Dataset & \#users & \#items & \#actions & avg. & density \\
         \hline
         Beauty  & 52,024 & 57,289 & 0.4M & 7.6 & 0.01\% \\
         Sports & 25,598 & 18,357 & 0.3M & 8.3  & 0.05\%\\
         Yelp &  30,431 & 20,033  & 0.3M & 10.4  & 0.05\%  \\
\bottomrule     
    \end{tabular}
    \label{tb1}
\end{table}
\begin{table*}[]
\caption{Model performance of backbones and our proposed UFNRec on three offline datasets, where `+UFN' refers to adding UFNRec to baselines, e.g., SASRec, BERT4Rec, SSE-PT.
\textit{Improv.} refers to the relative improvements of UFNRec over backbones, which are statistically significant with $p \textless 0.05$.}
\begin{tabular}{cllllllllll}
\toprule
Datesets  & Metrics & SASrec & +UFN   & Improv. & BERT4Rec & +UFN   & Improv. & SSE-PT & +UFN   & Improv. \\
\hline
\multirow{5}{*}{Beauty}        & HR@1    & 0.2040 & 0.2175 & 6.62\%  & 0.1919   & 0.1929 & 0.52\%  & 0.2013 & 0.2256 & 12.07\% \\
                               & HR@5    & 0.3785 & 0.3978 & 5.10\%  & 0.3689   & 0.3860 & 4.64\%  & 0.3884 & 0.4097 & 5.48\%  \\
                               & HR@10   & 0.4765 & 0.4953 & 3.95\%  & 0.4690   & 0.4877 & 3.99\%  & 0.4843 & 0.5002 & 3.28\%  \\
                               & NDCG@5  & 0.2942 & 0.3125 & 6.22\%  & 0.2864   & 0.2919 & 1.92\%  & 0.2990 & 0.3208 & 7.29\%  \\
                               & NDCG@10 & 0.3258 & 0.3439 & 5.56\%  & 0.3187   & 0.3247 & 1.88\%  & 0.3300 & 0.3500 & 6.06\%  \\
                               \hline
\multirow{5}{*}{Sports}        & HR@1    & 0.1797 & 0.1909 & 6.23\%  & 0.1619   & 0.1697 & 4.82\%  & 0.1856 & 0.1951 & 5.12\%  \\
                               & HR@5    & 0.3952 & 0.4060 & 2.73\%  & 0.3723   & 0.3841 & 3.17\%  & 0.3982 & 0.4029 & 1.18\%  \\
                               & HR@10   & 0.5072 & 0.5120 & 0.95\%  & 0.4936   & 0.5021 & 1.72\%  & 0.5073 & 0.5126 & 1.04\%  \\
                               & NDCG@5  & 0.2925 & 0.3049 & 4.24\%  & 0.2680   & 0.2775 & 3.54\%  & 0.2970 & 0.3024 & 1.82\%  \\
                               & NDCG@10 & 0.3286 & 0.3392 & 3.23\%  & 0.3071   & 0.3153 & 2.67\%  & 0.3322 & 0.3378 & 1.69\%  \\
                               \hline
\multirow{5}{*}{Yelp}          & HR@1    & 0.2743 & 0.2891 & 5.40\%  & 0.2686   & 0.2954 & 9.98\%  & 0.2979 & 0.3043 & 2.08\%  \\
                               & HR@5    & 0.6040 & 0.6239 & 3.29\%  & 0.6090   & 0.6475 & 6.32\%  & 0.6211 & 0.6351 & 2.22\%  \\
                               & HR@10   & 0.7589 & 0.7637 & 0.63\%  & 0.7530   & 0.7857 & 4.34\%  & 0.7554 & 0.7708 & 2.00\%  \\
                               & NDCG@5  & 0.4434 & 0.4681 & 5.57\%  & 0.4527   & 0.4829 & 6.67\%  & 0.4670 & 0.4773 & 2.15\%  \\
                               & NDCG@10 & 0.4934 & 0.5134 & 4.05\%  & 0.4993   & 0.5276 & 5.67\%  & 0.5106 & 0.5214 & 2.06\%  \\
                               \hline
\end{tabular}
\label{tb:main}
\end{table*}

\subsection{Baselines}
To prove the effectiveness of our proposed method, we choose three representative methods as baselines.

\begin{itemize}
 
\item \textbf{SASRec~\cite{sasrec}.} It utilizes the multi-head self-attention mechanism to tackle the SR task, and this model is widely regarded as one of the state-of-the-art baselines.

\item \textbf{BERT4Rec~\cite{BERT4Rec}.} 
It adopts a bidirectional self-attention mechanism to model user interaction sequences in the SR task. Like BERT~\cite{devlin2019bert}, this model predicts the masked items in the historical sequences during training. 

\item \textbf{SSE-PT~\cite{wu2020sse}.} 
SSE-PT proposes a personalized transformer architecture with a novel regularization technique of stochastic shared embeddings.
\end{itemize}

\subsection{Experiment Settings}
\noindent\textbf{Evaluation Metrics.}
We choose the leave-one-out strategy to evaluate model performance, which has been widely used in many previous studies~\cite{sasrec,CT4Rec}. Particularly, each user's last interacted item will be used for testing.
Similar to \cite{sasrec,s3}, for each positive item, we randomly sample 100 items from the whole dataset and rank them by prediction scores.
We evaluate model performance by HR\({@k}\) and NDCG\({@k}\) with $k=\left\{1,5,10\right\}$, which are commonly applied in SR tasks.

\noindent\textbf{Hyperparameter Settings.}
All models are implemented based on PyTorch with well-tested versions from the open-source community. We follow the original model settings of the correlated papers, i.e., the embedding dimension size.
All models are optimized by Adam with a learning rate of 0.001, and the batch size is set to 128. For all datasets, the maximum sequence length is 50.
We apply our UFNRec method to the above three backbone models by adding a false negative mining and reversing component (FMR) and a consistency training strategy (FCR). 
For the false negative reversing step, we conduct experiments with $m$ selected from $\{1,2,3,4,6,8,10\}$. To verify the influence of the loss weight $\alpha$ and the model decay rate $d$, we adjust the values of $\alpha$ from $\{0.01,0.05, 0.1,0.2,0.3,0.4,0.5\}$ and $d$ from $\{0.9, 0.99, 0.995, 0.999\}$. Besides, we test the influence of different batch sizes from $\{32, 64, 128, 256, 512\}$.
\subsection{Main Results(RQ1)}
As shown in Table~\ref{tb:main}, we report experimental results of all models on three real-world datasets to compare their performance in the sequential recommendation.
Notice that UFNRec only introduces a false negative mining and reversing component (FMR) and a consistency training objective (FCR) into backbones without involving extra data or changing the network structure. We compare \textit{UFNRec} with the three backbones to calibrate the influence of our method. 
We can observe that: 
1) \textit{SSE-PT} performs better than \textit{SASRec} and \textit{BERT4Rec} on most metrics, which indicates a personalized Transformer architecture is effective for the sequential recommendation. Furthermore, \textit{SSE-PT +UFN} can continuously improve the performance of \textit{SSE-PT}, achieving 2.10\% and 3.27\% (on average) improvements in terms of HR\({@10}\) and NDCG\({@10}\), respectively.
2) UFNRec improves the performance over three baselines on all metrics. 
Compared with the randomly sampling strategy applied in backbones, our UFNRec can mine false negative samples and utilize them to further improve model performance. 
The universal enhancements of \textit{UFNRec} over \textit{SASRec}, \textit{BERT4Rec}, and \textit{SSE-PT} on three datasets with all HR\({@k}\) and NDCG\({@k}\) scores (relative improvements ranging from 0.63\% to 12.07\%) generally verify the effectiveness of our model.
Thus, we can conclude that UFNRec is complementary to these backbones and can further improve the performance of SOTA models. 

\begin{table}[]
\caption{Ablation study of FMR and FCR on the Beauty dataset. `+FMR' refers to only adding the FMR operation to SASRec. `+FCR' refers to SASRec applying $\mathcal{L}_{final}$ without false negative reversing.}
\begin{tabular}{llllllll}
\toprule
Metrics & SASRec &  +FMR  &  +FCR  & +UFN  \\
\hline
HR@1	&	0.2040	&	0.2112	&	0.2081	&	0.2175	&	\\
HR@5	&	0.3785	&	0.3903	&	0.3882	&	0.3978	&	\\
HR@10	&	0.4765	&	0.4872	&	0.4836	&	0.4953	&	\\
NDCG@5	&	0.2942	&	0.3044	&	0.3025	&	0.3125	&	\\
NDCG@10	&	0.3258	&	0.3356	&	0.3333	&	0.3439	&	\\
\bottomrule
\end{tabular}
\label{tb_ablation}
\end{table}



\begin{table*}[]
\caption{Performance comparison of different strategies (i.e., removal vs. utilization) for false negative samples and different variants of UFNRec (i.e., +UFN$^{R}_{srns}$ and +UFN$_{srns}$) with varied false negative mining methods. Experiments are conducted on the Beauty dataset. Best performance is written in blod.}
\begin{tabular}{l|l|lll|lll}
\toprule
Strategies & origin & \multicolumn{3}{c|}{Removal}     & \multicolumn{3}{c}{Utilization} \\
\hline
Metrics  & SASRec & SASRec$^{R}$ & +SRNS$^{R}$ & +SRNS  & +UFN$^{R}_{srns}$ & +UFN$_{srns}$ & +UFN   \\
\hline
HR@1     & 0.2040  & 0.2119    & 0.2164     & 0.2062 & 0.2152      & \textbf{0.2197}   & 0.2175 \\
HR@5     & 0.3785 & 0.3886    & 0.3813     & 0.3879 & 0.3912      & 0.3864   &  \textbf{0.3978}  \\
HR@10    & 0.4765 & 0.4821    & 0.4744     & 0.4886 & 0.4897      & 0.4831   &  \textbf{0.4953} \\
NDCG@5   & 0.2942 & 0.3024    & 0.3028     & 0.3004 & 0.3073      & 0.3071   &  \textbf{0.3125} \\
NDCG@10  & 0.3258 & 0.3333    & 0.3329     & 0.3328 & 0.3385      & 0.3381   &  \textbf{0.3439} \\
\hline
\end{tabular}
\label{tb:SRNS}
\vspace{-0.3cm}
\end{table*}
\subsection{Ablation Study(RQ2)}
As illustrated in Table~\ref{tb_ablation}, we conduct an ablation study to explore the influence of two essential components mentioned in Section~\ref{sec:method}, i.e., the FMR and the FCR.

\noindent\textbf{False Negative Mining and Reversing(FMR).} As shown in Table~\ref{tb_ablation}, the FMR operation contributes significant improvements over the backbone \textit{SASRec} with a range from 1.87\% to 3.53\% on all metrics, which concludes that transferring false negative samples to positive samples can indeed improve model performance. In contrast, overlooking false negative samples can perturb the training process and yield inferior performance. 
Compared with the consistency regularization, the FMR operation contributes more improvements to the final results, which indicates that the FMR operation is more favorable to the model.
Also, we suppose that false negative mining is the foundation of UFNRec, and the firstly proposed label reversing operation is the core novelty of UFNRec.

\noindent\textbf{False Negative Consistency Regularization(FCR).} To only observe the effect of the consistency regularization objective, we maintain the false negative mining process and apply the consistency training objective to these samples without the label reversing step, which means false negative samples remain negative labels during the following training.
Table~\ref{tb_ablation} shows that the FCR objective can also yield performance improvements (relative improvements ranging from 1.49\% to 2.82\% on all metrics), which concludes that a consistency training objective to regularize the output distributions of false negative samples from the teacher-student framework is essential.
Furthermore, we can see that combining the FMR component and the FCR component can continuously enhance the model performance over a single part.

\begin{figure*}[]
    \centering
    \includegraphics[scale=0.42]{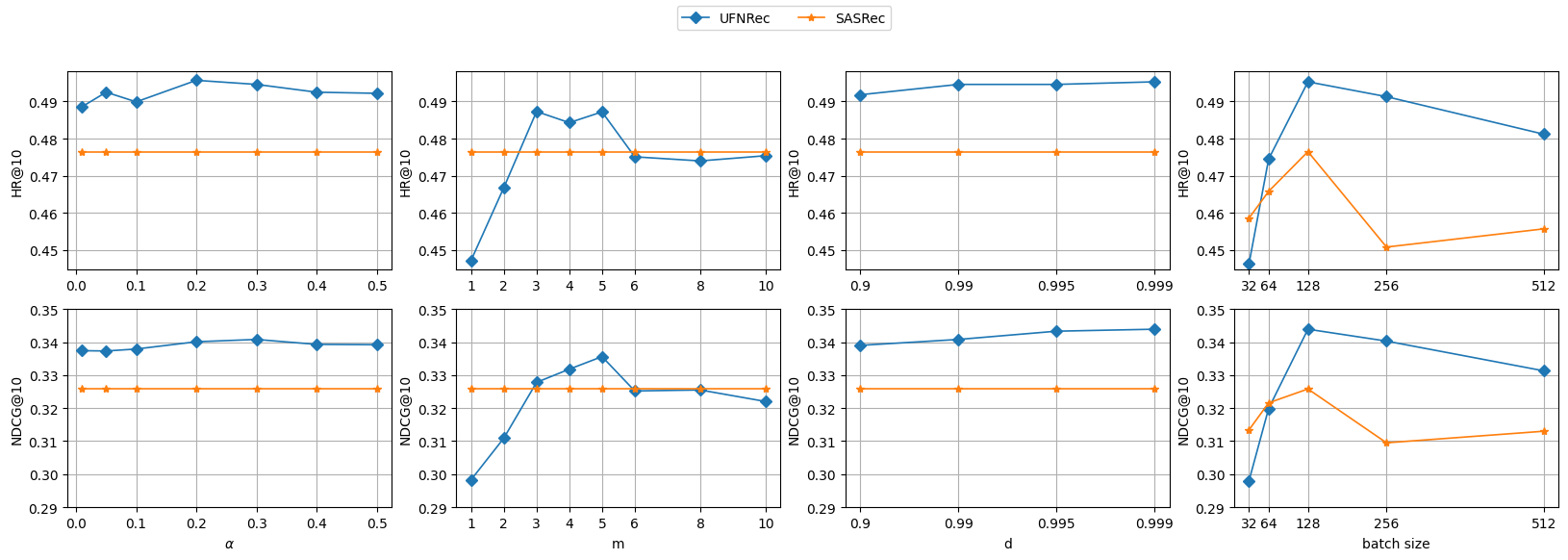}
    \caption{The impact of different hyper-parameters in terms of HR@10 and NDCG@10 for UFNRec and SASRec on the Beauty dataset. Hyper-parameters include the loss weight $\alpha$, the recorded times $m$, the decay rate $d$, and the batch size.}  \label{Fig:high_para}
    \vspace{-0.3cm}
    \end{figure*}
\subsection{Analysis on False Negative Mining and Removing(RQ3 and RQ4)}
In this part, we compare the newly proposed SRNS~\cite{ding2020simplify} with UFNRec.
SRNS introduced a variance-based negative sampling method and emphasized the effect of false negative samples. Different from UFNRec, SRNS focused on removing false negative samples to reduce their influence. 
Here, we treat SRNS as an additional method to mine false negative samples and compare the performance with UFNRec.
As shown in Table~\ref{tb:SRNS}, for false negative samples, we conduct experiments on \textit{Beauty} based on the backbone model \textit{SASRec} with three methods to remove false negative samples and three variants of UFNRec to utilize them. 
Specially, the application of SRNS can rely on $\mathcal{N}_{rec}$ in Section~\ref{sec:method_FMR} (i.e., \textit{+SRNS$^{R}$} and \textit{+UFN$_{srns}^{R}$}) or not (i.e., \textit{+SRNS} and \textit{+UFN$_{srns}$}).

Concretely, \textit{SASRec$^{R}$} utilizes the false negative mining method in Section~\ref{sec:method_FMR} and directly removes these samples during training.
As for \textit{+SRNS$^{R}$}, we apply the variance-based method proposed by SRNS~\cite{ding2020simplify} to distinguish false negative samples from the set $\mathcal{N}_{rec}$ and remove these samples.
Besides, we directly apply the variance-based method to mine false negative samples without relying on $\mathcal{N}_{rec}$ and remove these samples, which is denoted as \textit{+SRNS}.
Furthermore, for UFNRec, we apply SRNS to mine false negative samples and maintain all the remaining operations, including the label reversing step and the FCR part.
Similarly, we can filter out false negative samples based on $\mathcal{N}_{rec}$ (namely \textit{+UFN$_{srns}^{R}$}) or not (namely \textit{+UFN$_{srns}$}).

The performance comparisons of different strategies for false negative samples, including removal and utilization, are shown in Table~\ref{tb:SRNS}. 
Also, for UFNRec, we compare different variants of false negative mining methods. 
We can observe that: 1) The performance of \textit{SASRec} is worst on all metrics, which indicates overlooking false negative samples will lead to inferior model performance. 
2) Removal methods (e.g., \textit{+SASRec$^{R}$}, \textit{+SRNS$^{R}$}, and \textit{+SRNS}) can approximately perform better than \textit{SASRec} by reducing the effect of involving false negative samples.
3) Compared with simply removing these samples, \textit{+UFN} outperforms all removal methods in all evaluation metrics by adding a label reversing and a consistency training loss.
4) As for methods to utilize false negative samples,  \textit{+UFN} performs better than \textit{+UFN$_{srns}^{R}$} and \textit{+UFN$_{srns}$} on most evaluation metrics, which proves the effectiveness of our false negative mining method.

\subsection{Online A/B Test}
\textbf{Online Serving and Evaluation Protocol.}
We deploy UFNRec on a real-world recommendation platform used by over 300 million users. 
The online matching component consists of rule-based approaches and embedding-based approaches.
Concretely, we utilize UFNRec as an additional embedding-based retrieval approach in the matching component with other matching approaches unchanged. 
In the online experiment, we deploy UFNRec on two recommendation scenarios with two metrics including CVR and value per thousand impressions (VPM). We conduct the online A/B test for 5 days with 330 thousand users in the \textit{Tab Page} scenario and almost one million users in the \textit{First Page} scenario.
\begin{table}[]
\caption{Online A/B Tests.}
\begin{tabular}{lllll}
\toprule
Scenario  & CVR    & VPM  \\
\hline
Tab Page & +7.94\% & +5.83\%  \\
First Page & +8.60\% & +7.00\%   \\
\hline
\end{tabular}
\label{tb_online}
\vspace{-0.3cm}
\end{table}

\noindent\textbf{Experimental Results.}
Tabel~\ref{tb_online} shows the relative improvements of UFNRec, from which we can observe that: 
1) UFNRec achieves remarkable improvements on CVR and VPM metrics in both scenarios, which proves the effectiveness of UFNRec in online serving. 2) The \textit{Tab Page} is a small scenario with a few active users, where the corresponding user actions are relatively sparse and hard to learn. UFNRec can extend user interest from false negative samples, which enhances the recommendation performance.
3) The increment of CVR indicates that our system is attractive to more users. The improvement of VPM means that users attend to pay more in our system since UFNRec can draw user interests more accurately.

\subsection{Hyper-Parameter Analysis(RQ5)}
In this section, we mainly analyze the influence of four critical hyper-parameters for UFNRec, including the loss weight $\alpha$ in Equation~\ref{func:final_loss}, the recorded times $m$ to reverse labels, the decay rate $d$ in Equation~\ref{func:tea_update}, and the batch size.

\noindent\textbf{The Effect of $\alpha$.}
The first column in Figure~\ref{Fig:high_para} shows the performance of different $\alpha$ values. 
Since the combinations of the hyper-parameters are complicated, we temporarily fix all the other parameters to analyze the effect of $\alpha$.
As shown in Figure~\ref{Fig:high_para}, the consistency loss can achieve significant improvements with a small $\alpha$.
Both HR\({@10}\) and NDCG\({@10}\) slightly increase with the increase of $\alpha$ and then achieve the best result.
These experiment results further prove that the consistency regularization of the above teacher-student framework on the false negative samples is very powerful even when applying only a small $\alpha$ for the final loss during the training process.
With a continuous increase of $\alpha$, the HR\({@10}\) starts to decrease (e.g., $\alpha=0.2$ v.s.,$\alpha=0.5$) since the excessive emphasis on the consistency of outputs from the teacher-student framework can dilute the original objective and affect the model performance.

\noindent\textbf{The Effect of $m$.}
Similarly, we study the impact of different $m$ values with other parameters unchanged, as shown in the second column of Figure~\ref{Fig:high_para}.
Unlike $\alpha$, the reversing parameter $m$ leads to an inferior model performance when $m$\textless 3, which indicates that UFNRec with a small value of $m$ can not effectively distinguish false negative samples from negative samples and involve additional noise to the model.
With the increase of $m$, UFNRec gradually achieves a consistent improvement until $m$\textgreater 6. 
If $m$ is too large, and the model cannot discover false negative samples successfully.
The parameter $m$ is the benchmark to identify false negative samples from all negative samples. A small $m$ indicates a low criterion for false negative samples and vice versa.

\noindent\textbf{The Effect of $d$.}
The hyper-parameter of the EMA algorithm is the decay rate $d$. Here, we test the model performance's sensitivity to the parameter $d$. 
A smaller $d$ means the teacher model will quickly forget the old student models, and a larger $d$ means an extended memory teacher model. 
As shown in the third column of Figure~\ref{Fig:high_para}, we can see that the model performance is approximately stable, with $d$ from $0.9$ to $0.999$. 
Also, there is a slight increase in HR\({@10}\) and NDCG\({@10}\) when $d$ increases.
Since we warm up the model before the application of EMA, the student models improve and update relatively slowly. Thus, the model performance is not sensitive to the hyper-parameter $d$.
    
\noindent\textbf{The Effect of Batch Size.}
Here, we apply UFNRec to the backbone model SASRec and study the influence of different batch sizes ranging from 32 to 512. The fourth column in Figure~\ref{Fig:high_para} shows that UFNRec outperforms SASRec on HR\({@10}\) and NDCG\({@10}\) with a relatively large batch size (e.g., batch size \textgreater 64) on the Beauty dataset. We can see that though SASRec achieves the best performance at batch size=128, UFNRec can continuously improve the performance of SASRec.
Besides, UFNRec achieves a poorer performance than SASRec with a small batch size (i.e., 32), but a small batch size can constantly lead to a less reliable model and is rarely applied in online recommendation systems.
\begin{figure}[]
    \centering
    \includegraphics[scale=0.26]{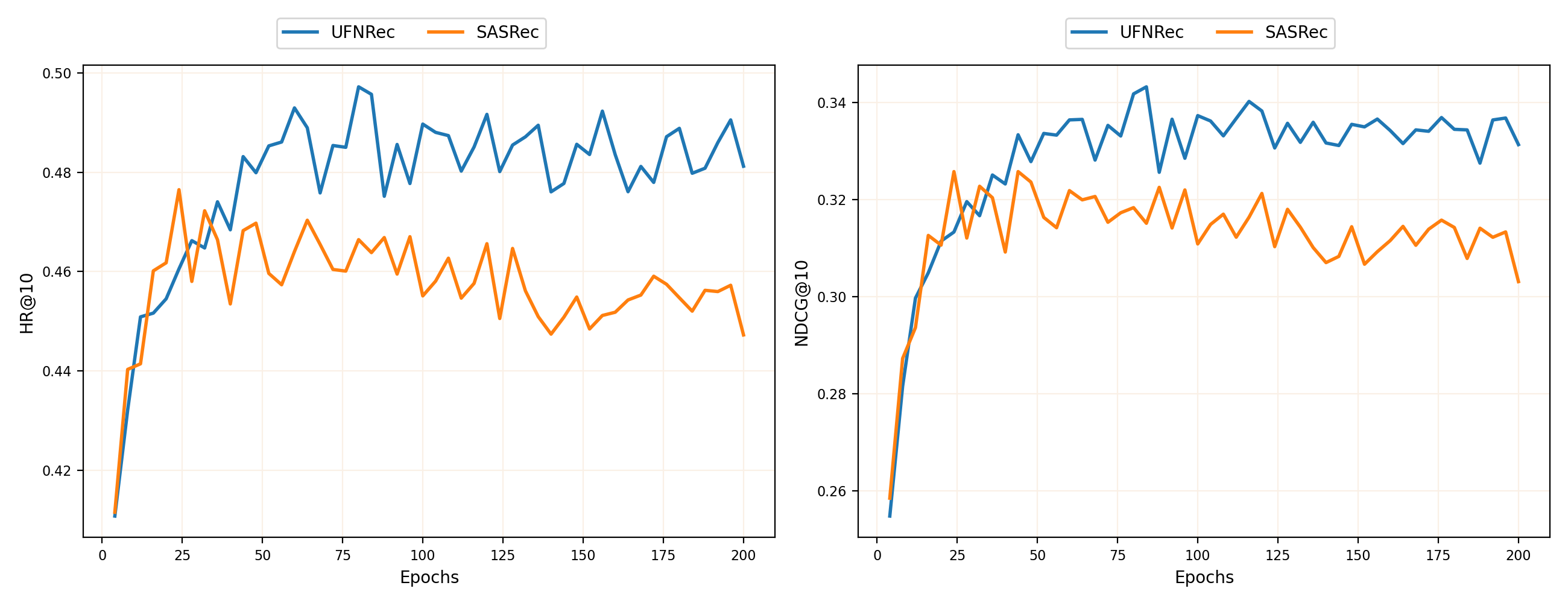}
    \caption{HR@10 and NDCG@10 curves of SASRec and UFNRec with varying training epochs on the Beauty dataset.}  
    \label{Fig:epoch}
    \vspace{-0.5cm}
    \end{figure} 

\begin{figure*}[t]
    \centering
    \includegraphics[scale=0.35]{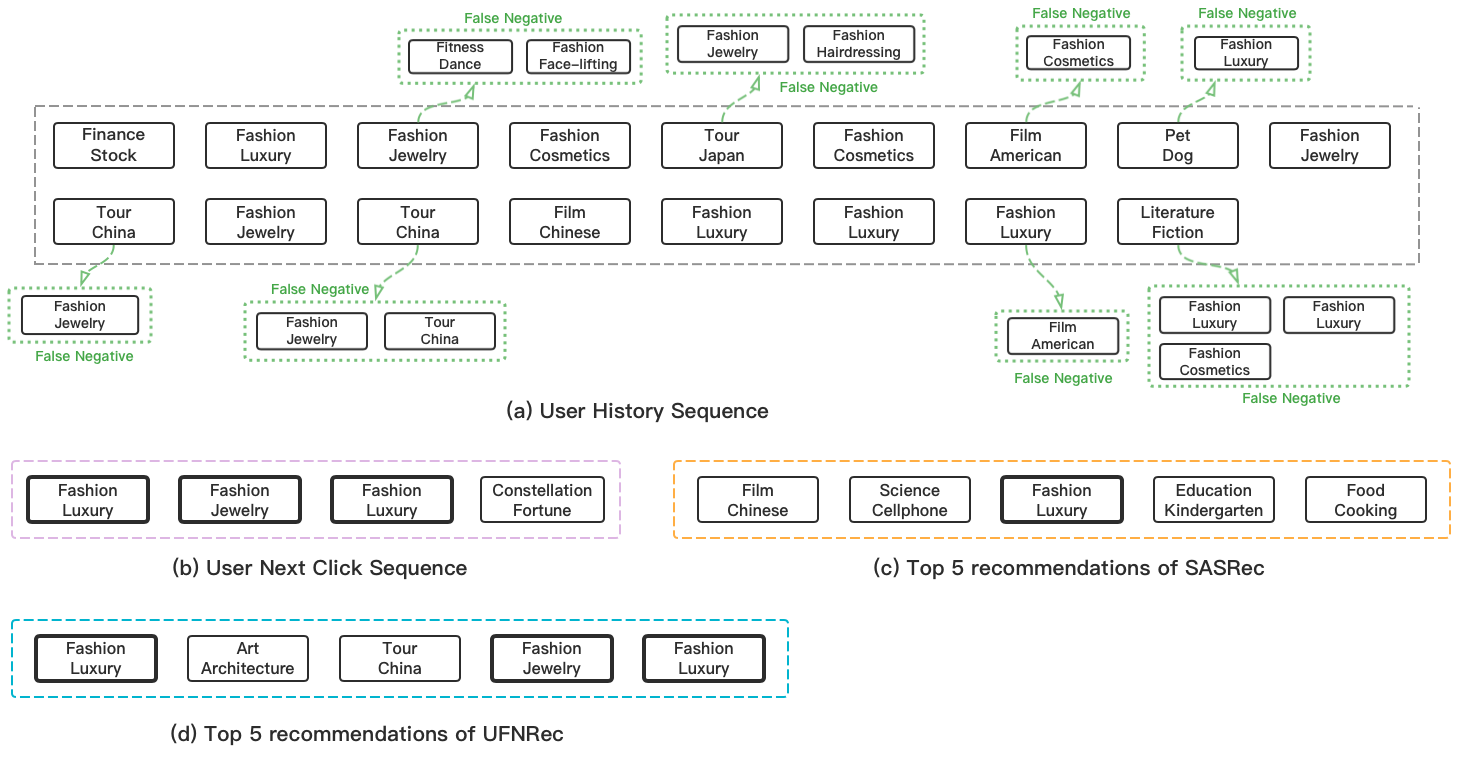}
    \caption{Case study on the sequential recommendation task. For simplicity, items are presented by corresponding categories. (a) A history sequence of a randomly sampled user from a real-world platform. Items in green boxes are false negative samples mined by UFNRec. (b) The next click sequence of the user following the history sequence. (c) and (d) are top 5 recommendation results of SASRec and UFNRec respectively. Items belonging to Fashion are shown in bold boxes.}  
    \label{Fig:example}
    \end{figure*} 
\subsection{Training Analysis(RQ6)}
Figure~\ref{Fig:epoch} shows the performance curves of SASRec and UFNRec with training epochs on the Beauty dataset. 
During the early training process, the performance of SASRec is slightly better than UFNRec with the same epochs (around 30 epochs) since our model needs some epochs to warm up the model and identify false negative samples.
After that, UFNRec continuously enhances the model performance and remarkably surpasses the well-trained SASRec.
Also, we can see that the performance of SASRec tends to decline with the increment of training epochs, while UFNRec can maintain the model performance at a high level and continuously outperform SASRec.
Here, we can conclude that UFNRec introduces additional training epochs to converge but can be alleviated by an early stop with the observation that UFNRec can outperform SASRec at early training epochs and achieve much better and stabler performance.

\subsection{Qualitative Analysis(RQ7)}
A crucial novelty of our model is that we realize the importance of false negative samples and utilize them to improve the sequential recommendation performance.
To better understand how UFNRec works, we qualitatively analyze a case from a real-world platform.
As shown at the top of Figure~\ref{Fig:example}, we randomly select one user and her history sequence containing seventeen clicked items.
For simplicity, we directly present the categories of these items, including major categories (i.e., finance and fashion) and detailed categories (i.e., stock and luxury).
Each item in the history sequence is a positive sample for the user, which is matched with randomly selected $100$ negative samples during the training process. 
Specially, items presented in green boxes are false negative samples mined by UFNRec from the corresponding $100$ negative samples.
Note that UFNRec distinguishes false negative samples from negative samples at each epoch. Here, we only present the false negative samples of this user at the last epoch.
Given the history sequence, we apply SASRec and UFNRec to recommend items that the user will click next.
For convenience, we present the top 5 recommendation results of SASRec and UFNRec in Figure~\ref{Fig:example}(c) and Figure~\ref{Fig:example}(d) to compare the model performance.
Also, Figure~\ref{Fig:example}(b) shows the next click sequence of the user, which is collected from the offline logs of the recommendation platform.

We can observe that the history sequence contains many items belonging to the fashion, which indicates the user has a strong interest in fashion.
Meanwhile, false negative samples that UFNRec discovers are mainly fashion items. Thus, directly training the model with these negative samples will disturb the training process.
As shown in Figure~\ref{Fig:example}, UFNRec recommends three fashion items while SASRec only recommends one.
Compared with the next click sequence of the user, the recommended items of UFNRec are better than SASRec and are more likely to be clicked by the user.
As we can see, UFNRec can discover false negative samples hidden in all negative samples and further utilize them to learn user interests.
In contrast, SASRec overlooks the existence of false negative samples and roughly trains the model with all randomly selected negative samples, which leads to a less robust model. 
Therefore, by utilizing false negative samples, our UFNRec can capture user interests more accurately and make the model more robust.

\section{Conclusion}
In this paper, we propose a simple yet effective method, named UFNRec, to utilize false negative samples for sequential recommendation tasks, which involves a false negative reversing step and a consistency training loss. 
We first design a strategy to distinguish false negative samples from true negative samples and transfer false negative samples to positive samples to train the model further.
Then, we introduce a teacher-student framework to provide soft labels for such false negative samples and then apply a consistency loss to regularize outputs from the framework.
Extensive experiments demonstrate the effectiveness and compatibility of our model.
To the best of our knowledge, this is the first work to utilize false negative samples to improve model performance for the sequential recommendation.
In the near future, we will explore more false negative mining methods for the recommendation systems.

\begin{acks}
We would like to thank the constructive suggestions of reviewers to improve this paper.
\end{acks}



\bibliographystyle{ACM-Reference-Format}
\bibliography{ref}


\begin{thebibliography}{75}


\ifx \showCODEN    \undefined \def \showCODEN     #1{\unskip}     \fi
\ifx \showDOI      \undefined \def \showDOI       #1{#1}\fi
\ifx \showISBNx    \undefined \def \showISBNx     #1{\unskip}     \fi
\ifx \showISBNxiii \undefined \def \showISBNxiii  #1{\unskip}     \fi
\ifx \showISSN     \undefined \def \showISSN      #1{\unskip}     \fi
\ifx \showLCCN     \undefined \def \showLCCN      #1{\unskip}     \fi
\ifx \shownote     \undefined \def \shownote      #1{#1}          \fi
\ifx \showarticletitle \undefined \def \showarticletitle #1{#1}   \fi
\ifx \showURL      \undefined \def \showURL       {\relax}        \fi
\providecommand\bibfield[2]{#2}
\providecommand\bibinfo[2]{#2}
\providecommand\natexlab[1]{#1}
\providecommand\showeprint[2][]{arXiv:#2}

\bibitem[\protect\citeauthoryear{Bengio and Senecal}{Bengio and
  Senecal}{2003}]%
        {pmlr-vR4-bengio03a}
\bibfield{author}{\bibinfo{person}{Yoshua Bengio} {and}
  \bibinfo{person}{Jean-S{\'{e}}bastien Senecal}.}
  \bibinfo{year}{2003}\natexlab{}.
\newblock \showarticletitle{Quick Training of Probabilistic Neural Nets by
  Importance Sampling}. In \bibinfo{booktitle}{\emph{Proceedings of the Ninth
  International Workshop on Artificial Intelligence and Statistics}}
  \emph{(\bibinfo{series}{Proceedings of Machine Learning Research},
  Vol.~\bibinfo{volume}{R4})},
  \bibfield{editor}{\bibinfo{person}{Christopher~M. Bishop} {and}
  \bibinfo{person}{Brendan~J. Frey}} (Eds.). \bibinfo{publisher}{PMLR},
  \bibinfo{pages}{17--24}.
\newblock


\bibitem[\protect\citeauthoryear{Bengio and Sen{\'e}cal}{Bengio and
  Sen{\'e}cal}{2008}]%
        {bengio2008adaptive}
\bibfield{author}{\bibinfo{person}{Yoshua Bengio} {and}
  \bibinfo{person}{Jean-S{\'e}bastien Sen{\'e}cal}.}
  \bibinfo{year}{2008}\natexlab{}.
\newblock \showarticletitle{Adaptive importance sampling to accelerate training
  of a neural probabilistic language model}.
\newblock \bibinfo{journal}{\emph{IEEE Transactions on Neural Networks}}
  \bibinfo{volume}{19}, \bibinfo{number}{4} (\bibinfo{year}{2008}),
  \bibinfo{pages}{713--722}.
\newblock


\bibitem[\protect\citeauthoryear{Beutel, Covington, Jain, Xu, Li, Gatto, and
  Chi}{Beutel et~al\mbox{.}}{2018}]%
        {Latent_Cross}
\bibfield{author}{\bibinfo{person}{Alex Beutel}, \bibinfo{person}{Paul
  Covington}, \bibinfo{person}{Sagar Jain}, \bibinfo{person}{Can Xu},
  \bibinfo{person}{Jia Li}, \bibinfo{person}{Vince Gatto}, {and}
  \bibinfo{person}{Ed~H Chi}.} \bibinfo{year}{2018}\natexlab{}.
\newblock \showarticletitle{Latent cross: Making use of context in recurrent
  recommender systems}. In \bibinfo{booktitle}{\emph{Proceedings of the
  Eleventh ACM International Conference on Web Search and Data Mining}}.
  \bibinfo{pages}{46--54}.
\newblock


\bibitem[\protect\citeauthoryear{Blanc and Rendle}{Blanc and Rendle}{2018}]%
        {blanc2018adaptive}
\bibfield{author}{\bibinfo{person}{Guy Blanc} {and} \bibinfo{person}{Steffen
  Rendle}.} \bibinfo{year}{2018}\natexlab{}.
\newblock \showarticletitle{Adaptive sampled softmax with kernel based
  sampling}. In \bibinfo{booktitle}{\emph{International Conference on Machine
  Learning}}. PMLR, \bibinfo{pages}{590--599}.
\newblock


\bibitem[\protect\citeauthoryear{Chen, Sun, Shi, and Hong}{Chen
  et~al\mbox{.}}{2017}]%
        {chen2017sampling}
\bibfield{author}{\bibinfo{person}{Ting Chen}, \bibinfo{person}{Yizhou Sun},
  \bibinfo{person}{Yue Shi}, {and} \bibinfo{person}{Liangjie Hong}.}
  \bibinfo{year}{2017}\natexlab{}.
\newblock \showarticletitle{On sampling strategies for neural network-based
  collaborative filtering}. In \bibinfo{booktitle}{\emph{Proceedings of the
  23rd ACM SIGKDD International Conference on Knowledge Discovery and Data
  Mining}}. \bibinfo{pages}{767--776}.
\newblock


\bibitem[\protect\citeauthoryear{Chen, Yin, Chen, Yan, Nguyen, and Li}{Chen
  et~al\mbox{.}}{2019}]%
        {Air}
\bibfield{author}{\bibinfo{person}{Tong Chen}, \bibinfo{person}{Hongzhi Yin},
  \bibinfo{person}{Hongxu Chen}, \bibinfo{person}{Rui Yan},
  \bibinfo{person}{Quoc Viet~Hung Nguyen}, {and} \bibinfo{person}{Xue Li}.}
  \bibinfo{year}{2019}\natexlab{}.
\newblock \showarticletitle{Air: Attentional intention-aware recommender
  systems}. In \bibinfo{booktitle}{\emph{2019 IEEE 35th International
  Conference on Data Engineering (ICDE)}}. IEEE, \bibinfo{pages}{304--315}.
\newblock


\bibitem[\protect\citeauthoryear{Chen, Xu, Zhang, Tang, Cao, Qin, and Zha}{Chen
  et~al\mbox{.}}{2018}]%
        {MANN}
\bibfield{author}{\bibinfo{person}{Xu Chen}, \bibinfo{person}{Hongteng Xu},
  \bibinfo{person}{Yongfeng Zhang}, \bibinfo{person}{Jiaxi Tang},
  \bibinfo{person}{Yixin Cao}, \bibinfo{person}{Zheng Qin}, {and}
  \bibinfo{person}{Hongyuan Zha}.} \bibinfo{year}{2018}\natexlab{}.
\newblock \showarticletitle{Sequential recommendation with user memory
  networks}. In \bibinfo{booktitle}{\emph{Proceedings of the eleventh ACM
  international conference on web search and data mining}}.
  \bibinfo{pages}{108--116}.
\newblock


\bibitem[\protect\citeauthoryear{Covington, Adams, and Sargin}{Covington
  et~al\mbox{.}}{2016}]%
        {youtubeDNN}
\bibfield{author}{\bibinfo{person}{Paul Covington}, \bibinfo{person}{Jay
  Adams}, {and} \bibinfo{person}{Emre Sargin}.}
  \bibinfo{year}{2016}\natexlab{}.
\newblock \showarticletitle{Deep neural networks for youtube recommendations}.
  In \bibinfo{booktitle}{\emph{Proceedings of the 10th ACM conference on
  recommender systems}}. \bibinfo{pages}{191--198}.
\newblock


\bibitem[\protect\citeauthoryear{Devlin, Chang, Lee, and Toutanova}{Devlin
  et~al\mbox{.}}{2019}]%
        {devlin2019bert}
\bibfield{author}{\bibinfo{person}{Jacob Devlin}, \bibinfo{person}{Ming-Wei
  Chang}, \bibinfo{person}{Kenton Lee}, {and} \bibinfo{person}{Kristina
  Toutanova}.} \bibinfo{year}{2019}\natexlab{}.
\newblock \showarticletitle{BERT: Pre-training of Deep Bidirectional
  Transformers for Language Understanding}. In
  \bibinfo{booktitle}{\emph{Proceedings of the 2019 Conference of the North
  American Chapter of the Association for Computational Linguistics: Human
  Language Technologies, Volume 1 (Long and Short Papers)}}.
  \bibinfo{pages}{4171--4186}.
\newblock


\bibitem[\protect\citeauthoryear{Ding, Quan, Yao, Li, and Jin}{Ding
  et~al\mbox{.}}{2020}]%
        {ding2020simplify}
\bibfield{author}{\bibinfo{person}{Jingtao Ding}, \bibinfo{person}{Yuhan Quan},
  \bibinfo{person}{Quanming Yao}, \bibinfo{person}{Yong Li}, {and}
  \bibinfo{person}{Depeng Jin}.} \bibinfo{year}{2020}\natexlab{}.
\newblock \showarticletitle{Simplify and robustify negative sampling for
  implicit collaborative filtering}.
\newblock \bibinfo{journal}{\emph{Advances in Neural Information Processing
  Systems}}  \bibinfo{volume}{33} (\bibinfo{year}{2020}),
  \bibinfo{pages}{1094--1105}.
\newblock


\bibitem[\protect\citeauthoryear{Donkers, Loepp, and Ziegler}{Donkers
  et~al\mbox{.}}{2017}]%
        {UserBased}
\bibfield{author}{\bibinfo{person}{Tim Donkers}, \bibinfo{person}{Benedikt
  Loepp}, {and} \bibinfo{person}{J{\"u}rgen Ziegler}.}
  \bibinfo{year}{2017}\natexlab{}.
\newblock \showarticletitle{Sequential user-based recurrent neural network
  recommendations}. In \bibinfo{booktitle}{\emph{Proceedings of the eleventh
  ACM conference on recommender systems}}. \bibinfo{pages}{152--160}.
\newblock


\bibitem[\protect\citeauthoryear{Fan, Liu, Lian, Zhao, Xie, and Wen}{Fan
  et~al\mbox{.}}{2021}]%
        {fan2021lighter}
\bibfield{author}{\bibinfo{person}{Xinyan Fan}, \bibinfo{person}{Zheng Liu},
  \bibinfo{person}{Jianxun Lian}, \bibinfo{person}{Wayne~Xin Zhao},
  \bibinfo{person}{Xing Xie}, {and} \bibinfo{person}{Ji-Rong Wen}.}
  \bibinfo{year}{2021}\natexlab{}.
\newblock \showarticletitle{Lighter and better: low-rank decomposed
  self-attention networks for next-item recommendation}. In
  \bibinfo{booktitle}{\emph{Proceedings of the 44th International ACM SIGIR
  Conference on Research and Development in Information Retrieval}}.
  \bibinfo{pages}{1733--1737}.
\newblock


\bibitem[\protect\citeauthoryear{Gillick, Kulkarni, Lansing, Presta, Baldridge,
  Ie, and Garcia-Olano}{Gillick et~al\mbox{.}}{2019}]%
        {hard-1}
\bibfield{author}{\bibinfo{person}{Daniel Gillick}, \bibinfo{person}{Sayali
  Kulkarni}, \bibinfo{person}{Larry Lansing}, \bibinfo{person}{Alessandro
  Presta}, \bibinfo{person}{Jason Baldridge}, \bibinfo{person}{Eugene Ie},
  {and} \bibinfo{person}{Diego Garcia-Olano}.} \bibinfo{year}{2019}\natexlab{}.
\newblock \showarticletitle{Learning dense representations for entity
  retrieval}.
\newblock \bibinfo{journal}{\emph{arXiv preprint arXiv:1909.10506}}
  (\bibinfo{year}{2019}).
\newblock


\bibitem[\protect\citeauthoryear{Guo, Mousavi, Wu, Holtmann-Rice, Kale, Reddi,
  and Kumar}{Guo et~al\mbox{.}}{2019}]%
        {GLaS}
\bibfield{author}{\bibinfo{person}{Chuan Guo}, \bibinfo{person}{Ali Mousavi},
  \bibinfo{person}{Xiang Wu}, \bibinfo{person}{Dan Holtmann-Rice},
  \bibinfo{person}{Satyen Kale}, \bibinfo{person}{Sashank Reddi}, {and}
  \bibinfo{person}{Sanjiv Kumar}.} \bibinfo{year}{2019}\natexlab{}.
\newblock \showarticletitle{Breaking the glass ceiling for embedding-based
  classifiers for large output spaces}.
\newblock  (\bibinfo{year}{2019}).
\newblock


\bibitem[\protect\citeauthoryear{He, Kang, and McAuley}{He
  et~al\mbox{.}}{2017}]%
        {TransRec}
\bibfield{author}{\bibinfo{person}{Ruining He}, \bibinfo{person}{Wang-Cheng
  Kang}, {and} \bibinfo{person}{Julian McAuley}.}
  \bibinfo{year}{2017}\natexlab{}.
\newblock \showarticletitle{Translation-based recommendation}. In
  \bibinfo{booktitle}{\emph{Proceedings of the eleventh ACM conference on
  recommender systems}}. \bibinfo{pages}{161--169}.
\newblock


\bibitem[\protect\citeauthoryear{He and McAuley}{He and McAuley}{2016}]%
        {Fossil}
\bibfield{author}{\bibinfo{person}{Ruining He} {and} \bibinfo{person}{Julian
  McAuley}.} \bibinfo{year}{2016}\natexlab{}.
\newblock \showarticletitle{Fusing similarity models with markov chains for
  sparse sequential recommendation}. In \bibinfo{booktitle}{\emph{2016 IEEE
  16th International Conference on Data Mining (ICDM)}}. IEEE,
  \bibinfo{pages}{191--200}.
\newblock


\bibitem[\protect\citeauthoryear{Hern{\'a}ndez-Lobato, Houlsby, and
  Ghahramani}{Hern{\'a}ndez-Lobato et~al\mbox{.}}{2014}]%
        {hernandez2014probabilistic}
\bibfield{author}{\bibinfo{person}{Jos{\'e}~Miguel Hern{\'a}ndez-Lobato},
  \bibinfo{person}{Neil Houlsby}, {and} \bibinfo{person}{Zoubin Ghahramani}.}
  \bibinfo{year}{2014}\natexlab{}.
\newblock \showarticletitle{Probabilistic matrix factorization with non-random
  missing data}. In \bibinfo{booktitle}{\emph{International Conference on
  Machine Learning}}. PMLR, \bibinfo{pages}{1512--1520}.
\newblock


\bibitem[\protect\citeauthoryear{Hidasi and Karatzoglou}{Hidasi and
  Karatzoglou}{2018}]%
        {lossRNN}
\bibfield{author}{\bibinfo{person}{Bal{\'a}zs Hidasi} {and}
  \bibinfo{person}{Alexandros Karatzoglou}.} \bibinfo{year}{2018}\natexlab{}.
\newblock \showarticletitle{Recurrent neural networks with top-k gains for
  session-based recommendations}. In \bibinfo{booktitle}{\emph{Proceedings of
  the 27th ACM international conference on information and knowledge
  management}}. \bibinfo{pages}{843--852}.
\newblock


\bibitem[\protect\citeauthoryear{Hidasi, Karatzoglou, Baltrunas, and
  Tikk}{Hidasi et~al\mbox{.}}{2015}]%
        {GRU4Rec}
\bibfield{author}{\bibinfo{person}{Bal{\'a}zs Hidasi},
  \bibinfo{person}{Alexandros Karatzoglou}, \bibinfo{person}{Linas Baltrunas},
  {and} \bibinfo{person}{Domonkos Tikk}.} \bibinfo{year}{2015}\natexlab{}.
\newblock \showarticletitle{Session-based recommendations with recurrent neural
  networks}.
\newblock \bibinfo{journal}{\emph{arXiv preprint arXiv:1511.06939}}
  (\bibinfo{year}{2015}).
\newblock


\bibitem[\protect\citeauthoryear{Hochreiter and Schmidhuber}{Hochreiter and
  Schmidhuber}{1997}]%
        {LSTM}
\bibfield{author}{\bibinfo{person}{Sepp Hochreiter} {and}
  \bibinfo{person}{J{\"u}rgen Schmidhuber}.} \bibinfo{year}{1997}\natexlab{}.
\newblock \showarticletitle{Long short-term memory}.
\newblock \bibinfo{journal}{\emph{Neural computation}} \bibinfo{volume}{9},
  \bibinfo{number}{8} (\bibinfo{year}{1997}), \bibinfo{pages}{1735--1780}.
\newblock


\bibitem[\protect\citeauthoryear{Hu, Cao, Wang, Xu, Cao, and Gu}{Hu
  et~al\mbox{.}}{2017}]%
        {2017hu}
\bibfield{author}{\bibinfo{person}{Liang Hu}, \bibinfo{person}{Longbing Cao},
  \bibinfo{person}{Shoujin Wang}, \bibinfo{person}{Guandong Xu},
  \bibinfo{person}{Jian Cao}, {and} \bibinfo{person}{Zhiping Gu}.}
  \bibinfo{year}{2017}\natexlab{}.
\newblock \showarticletitle{Diversifying Personalized Recommendation with
  User-session Context.}. In \bibinfo{booktitle}{\emph{IJCAI}}.
  \bibinfo{pages}{1858--1864}.
\newblock


\bibitem[\protect\citeauthoryear{Huang, Zhao, Dou, Wen, and Chang}{Huang
  et~al\mbox{.}}{2018}]%
        {KV-MN}
\bibfield{author}{\bibinfo{person}{Jin Huang}, \bibinfo{person}{Wayne~Xin
  Zhao}, \bibinfo{person}{Hongjian Dou}, \bibinfo{person}{Ji-Rong Wen}, {and}
  \bibinfo{person}{Edward~Y Chang}.} \bibinfo{year}{2018}\natexlab{}.
\newblock \showarticletitle{Improving sequential recommendation with
  knowledge-enhanced memory networks}. In \bibinfo{booktitle}{\emph{The 41st
  International ACM SIGIR Conference on Research \& Development in Information
  Retrieval}}. \bibinfo{pages}{505--514}.
\newblock


\bibitem[\protect\citeauthoryear{Huang, Sharma, Sun, Xia, Zhang, Pronin,
  Padmanabhan, Ottaviano, and Yang}{Huang et~al\mbox{.}}{2020}]%
        {ebr}
\bibfield{author}{\bibinfo{person}{Jui-Ting Huang}, \bibinfo{person}{Ashish
  Sharma}, \bibinfo{person}{Shuying Sun}, \bibinfo{person}{Li Xia},
  \bibinfo{person}{David Zhang}, \bibinfo{person}{Philip Pronin},
  \bibinfo{person}{Janani Padmanabhan}, \bibinfo{person}{Giuseppe Ottaviano},
  {and} \bibinfo{person}{Linjun Yang}.} \bibinfo{year}{2020}\natexlab{}.
\newblock \showarticletitle{Embedding-based retrieval in facebook search}. In
  \bibinfo{booktitle}{\emph{Proceedings of the 26th ACM SIGKDD International
  Conference on Knowledge Discovery \& Data Mining}}.
  \bibinfo{pages}{2553--2561}.
\newblock


\bibitem[\protect\citeauthoryear{Jing and Smola}{Jing and Smola}{2017}]%
        {jing2017neural}
\bibfield{author}{\bibinfo{person}{How Jing} {and} \bibinfo{person}{Alexander~J
  Smola}.} \bibinfo{year}{2017}\natexlab{}.
\newblock \showarticletitle{Neural survival recommender}. In
  \bibinfo{booktitle}{\emph{Proceedings of the Tenth ACM International
  Conference on Web Search and Data Mining}}. \bibinfo{pages}{515--524}.
\newblock


\bibitem[\protect\citeauthoryear{Kang, Lee, Choe, and Jung}{Kang
  et~al\mbox{.}}{2021}]%
        {E-BART4Rec}
\bibfield{author}{\bibinfo{person}{Taegwan Kang}, \bibinfo{person}{Hwanhee
  Lee}, \bibinfo{person}{Byeongjin Choe}, {and} \bibinfo{person}{Kyomin Jung}.}
  \bibinfo{year}{2021}\natexlab{}.
\newblock \showarticletitle{Entangled Bidirectional Encoder to Autoregressive
  Decoder for Sequential Recommendation}. In
  \bibinfo{booktitle}{\emph{Proceedings of the 44th International ACM SIGIR
  Conference on Research and Development in Information Retrieval}}.
  \bibinfo{pages}{1657--1661}.
\newblock


\bibitem[\protect\citeauthoryear{Kang and McAuley}{Kang and McAuley}{2018}]%
        {sasrec}
\bibfield{author}{\bibinfo{person}{Wang-Cheng Kang} {and}
  \bibinfo{person}{Julian McAuley}.} \bibinfo{year}{2018}\natexlab{}.
\newblock \showarticletitle{Self-attentive sequential recommendation}. In
  \bibinfo{booktitle}{\emph{2018 IEEE International Conference on Data Mining
  (ICDM)}}. IEEE, \bibinfo{pages}{197--206}.
\newblock


\bibitem[\protect\citeauthoryear{Karpukhin, O{\u{g}}uz, Min, Lewis, Wu, Edunov,
  Chen, and Yih}{Karpukhin et~al\mbox{.}}{2020}]%
        {hard-2}
\bibfield{author}{\bibinfo{person}{Vladimir Karpukhin}, \bibinfo{person}{Barlas
  O{\u{g}}uz}, \bibinfo{person}{Sewon Min}, \bibinfo{person}{Patrick Lewis},
  \bibinfo{person}{Ledell Wu}, \bibinfo{person}{Sergey Edunov},
  \bibinfo{person}{Danqi Chen}, {and} \bibinfo{person}{Wen-tau Yih}.}
  \bibinfo{year}{2020}\natexlab{}.
\newblock \showarticletitle{Dense passage retrieval for open-domain question
  answering}.
\newblock \bibinfo{journal}{\emph{arXiv preprint arXiv:2004.04906}}
  (\bibinfo{year}{2020}).
\newblock


\bibitem[\protect\citeauthoryear{Kipf and Welling}{Kipf and Welling}{2016}]%
        {GCN}
\bibfield{author}{\bibinfo{person}{Thomas~N Kipf} {and} \bibinfo{person}{Max
  Welling}.} \bibinfo{year}{2016}\natexlab{}.
\newblock \showarticletitle{Semi-supervised classification with graph
  convolutional networks}.
\newblock \bibinfo{journal}{\emph{arXiv preprint arXiv:1609.02907}}
  (\bibinfo{year}{2016}).
\newblock


\bibitem[\protect\citeauthoryear{Li, Ren, Chen, Ren, Lian, and Ma}{Li
  et~al\mbox{.}}{2017}]%
        {NARM}
\bibfield{author}{\bibinfo{person}{Jing Li}, \bibinfo{person}{Pengjie Ren},
  \bibinfo{person}{Zhumin Chen}, \bibinfo{person}{Zhaochun Ren},
  \bibinfo{person}{Tao Lian}, {and} \bibinfo{person}{Jun Ma}.}
  \bibinfo{year}{2017}\natexlab{}.
\newblock \showarticletitle{Neural attentive session-based recommendation}. In
  \bibinfo{booktitle}{\emph{Proceedings of the 2017 ACM on Conference on
  Information and Knowledge Management}}. \bibinfo{pages}{1419--1428}.
\newblock


\bibitem[\protect\citeauthoryear{Li, Wang, and McAuley}{Li
  et~al\mbox{.}}{2020}]%
        {TiSASRec}
\bibfield{author}{\bibinfo{person}{Jiacheng Li}, \bibinfo{person}{Yujie Wang},
  {and} \bibinfo{person}{Julian McAuley}.} \bibinfo{year}{2020}\natexlab{}.
\newblock \showarticletitle{Time interval aware self-attention for sequential
  recommendation}. In \bibinfo{booktitle}{\emph{Proceedings of the 13th
  international conference on web search and data mining}}.
  \bibinfo{pages}{322--330}.
\newblock


\bibitem[\protect\citeauthoryear{Liang, Charlin, McInerney, and Blei}{Liang
  et~al\mbox{.}}{2016}]%
        {liang2016modeling}
\bibfield{author}{\bibinfo{person}{Dawen Liang}, \bibinfo{person}{Laurent
  Charlin}, \bibinfo{person}{James McInerney}, {and} \bibinfo{person}{David~M
  Blei}.} \bibinfo{year}{2016}\natexlab{}.
\newblock \showarticletitle{Modeling user exposure in recommendation}. In
  \bibinfo{booktitle}{\emph{Proceedings of the 25th international conference on
  World Wide Web}}. \bibinfo{pages}{951--961}.
\newblock


\bibitem[\protect\citeauthoryear{Linden, Smith, and York}{Linden
  et~al\mbox{.}}{2003}]%
        {amazon}
\bibfield{author}{\bibinfo{person}{Greg Linden}, \bibinfo{person}{Brent Smith},
  {and} \bibinfo{person}{Jeremy York}.} \bibinfo{year}{2003}\natexlab{}.
\newblock \showarticletitle{Amazon. com recommendations: Item-to-item
  collaborative filtering}.
\newblock \bibinfo{journal}{\emph{IEEE Internet computing}}
  \bibinfo{volume}{7}, \bibinfo{number}{1} (\bibinfo{year}{2003}),
  \bibinfo{pages}{76--80}.
\newblock


\bibitem[\protect\citeauthoryear{Liu, Liu, Zheng, Zhang, Liang, Li, Wu, Zhang,
  and Lin}{Liu et~al\mbox{.}}{2021b}]%
        {CT4Rec}
\bibfield{author}{\bibinfo{person}{Chong Liu}, \bibinfo{person}{Xiaoyang Liu},
  \bibinfo{person}{Rongqin Zheng}, \bibinfo{person}{Lixin Zhang},
  \bibinfo{person}{Xiaobo Liang}, \bibinfo{person}{Juntao Li},
  \bibinfo{person}{Lijun Wu}, \bibinfo{person}{Min Zhang}, {and}
  \bibinfo{person}{Leyu Lin}.} \bibinfo{year}{2021}\natexlab{b}.
\newblock \showarticletitle{C$^{2}$-Rec: An Effective Consistency Constraint
  for Sequential Recommendation}.
\newblock \bibinfo{journal}{\emph{arXiv preprint arXiv:2112.06668}}
  (\bibinfo{year}{2021}).
\newblock


\bibitem[\protect\citeauthoryear{Liu, Wu, Wang, Li, and Wang}{Liu
  et~al\mbox{.}}{2016}]%
        {liu2016context}
\bibfield{author}{\bibinfo{person}{Qiang Liu}, \bibinfo{person}{Shu Wu},
  \bibinfo{person}{Diyi Wang}, \bibinfo{person}{Zhaokang Li}, {and}
  \bibinfo{person}{Liang Wang}.} \bibinfo{year}{2016}\natexlab{}.
\newblock \showarticletitle{Context-aware sequential recommendation}. In
  \bibinfo{booktitle}{\emph{2016 IEEE 16th International Conference on Data
  Mining (ICDM)}}. IEEE, \bibinfo{pages}{1053--1058}.
\newblock


\bibitem[\protect\citeauthoryear{Liu, Fan, Wang, and Yu}{Liu
  et~al\mbox{.}}{2021a}]%
        {liu2021augmenting}
\bibfield{author}{\bibinfo{person}{Zhiwei Liu}, \bibinfo{person}{Ziwei Fan},
  \bibinfo{person}{Yu Wang}, {and} \bibinfo{person}{Philip~S Yu}.}
  \bibinfo{year}{2021}\natexlab{a}.
\newblock \showarticletitle{Augmenting Sequential Recommendation with
  Pseudo-Prior Items via Reversely Pre-training Transformer}.
\newblock \bibinfo{journal}{\emph{arXiv preprint arXiv:2105.00522}}
  (\bibinfo{year}{2021}).
\newblock


\bibitem[\protect\citeauthoryear{Lv, Jin, Yu, Sun, Lin, Yang, and Ng}{Lv
  et~al\mbox{.}}{2019}]%
        {SDM}
\bibfield{author}{\bibinfo{person}{Fuyu Lv}, \bibinfo{person}{Taiwei Jin},
  \bibinfo{person}{Changlong Yu}, \bibinfo{person}{Fei Sun},
  \bibinfo{person}{Quan Lin}, \bibinfo{person}{Keping Yang}, {and}
  \bibinfo{person}{Wilfred Ng}.} \bibinfo{year}{2019}\natexlab{}.
\newblock \showarticletitle{SDM: Sequential deep matching model for online
  large-scale recommender system}. In \bibinfo{booktitle}{\emph{Proceedings of
  the 28th ACM International Conference on Information and Knowledge
  Management}}. \bibinfo{pages}{2635--2643}.
\newblock


\bibitem[\protect\citeauthoryear{Ma, Zhou, Yang, Cui, Wang, and Zhu}{Ma
  et~al\mbox{.}}{2020}]%
        {disentangled1}
\bibfield{author}{\bibinfo{person}{Jianxin Ma}, \bibinfo{person}{Chang Zhou},
  \bibinfo{person}{Hongxia Yang}, \bibinfo{person}{Peng Cui},
  \bibinfo{person}{Xin Wang}, {and} \bibinfo{person}{Wenwu Zhu}.}
  \bibinfo{year}{2020}\natexlab{}.
\newblock \showarticletitle{Disentangled self-supervision in sequential
  recommenders}. In \bibinfo{booktitle}{\emph{Proceedings of the 26th ACM
  SIGKDD International Conference on Knowledge Discovery \& Data Mining}}.
  \bibinfo{pages}{483--491}.
\newblock


\bibitem[\protect\citeauthoryear{McAuley, Targett, Shi, and Van
  Den~Hengel}{McAuley et~al\mbox{.}}{2015}]%
        {amazon_datasets}
\bibfield{author}{\bibinfo{person}{Julian McAuley},
  \bibinfo{person}{Christopher Targett}, \bibinfo{person}{Qinfeng Shi}, {and}
  \bibinfo{person}{Anton Van Den~Hengel}.} \bibinfo{year}{2015}\natexlab{}.
\newblock \showarticletitle{Image-based recommendations on styles and
  substitutes}. In \bibinfo{booktitle}{\emph{Proceedings of the 38th
  international ACM SIGIR conference on research and development in information
  retrieval}}. \bibinfo{pages}{43--52}.
\newblock


\bibitem[\protect\citeauthoryear{Mikolov, Sutskever, Chen, Corrado, and
  Dean}{Mikolov et~al\mbox{.}}{2013}]%
        {mikolov2013distributed}
\bibfield{author}{\bibinfo{person}{Tomas Mikolov}, \bibinfo{person}{Ilya
  Sutskever}, \bibinfo{person}{Kai Chen}, \bibinfo{person}{Greg~S Corrado},
  {and} \bibinfo{person}{Jeff Dean}.} \bibinfo{year}{2013}\natexlab{}.
\newblock \showarticletitle{Distributed representations of words and phrases
  and their compositionality}. In \bibinfo{booktitle}{\emph{Advances in neural
  information processing systems}}. \bibinfo{pages}{3111--3119}.
\newblock


\bibitem[\protect\citeauthoryear{Qiu, Li, Huang, and Yin}{Qiu
  et~al\mbox{.}}{2019}]%
        {FGNN}
\bibfield{author}{\bibinfo{person}{Ruihong Qiu}, \bibinfo{person}{Jingjing Li},
  \bibinfo{person}{Zi Huang}, {and} \bibinfo{person}{Hongzhi Yin}.}
  \bibinfo{year}{2019}\natexlab{}.
\newblock \showarticletitle{Rethinking the item order in session-based
  recommendation with graph neural networks}. In
  \bibinfo{booktitle}{\emph{Proceedings of the 28th ACM International
  Conference on Information and Knowledge Management}}.
  \bibinfo{pages}{579--588}.
\newblock


\bibitem[\protect\citeauthoryear{Quadrana, Karatzoglou, Hidasi, and
  Cremonesi}{Quadrana et~al\mbox{.}}{2017}]%
        {HRNN}
\bibfield{author}{\bibinfo{person}{Massimo Quadrana},
  \bibinfo{person}{Alexandros Karatzoglou}, \bibinfo{person}{Bal{\'a}zs
  Hidasi}, {and} \bibinfo{person}{Paolo Cremonesi}.}
  \bibinfo{year}{2017}\natexlab{}.
\newblock \showarticletitle{Personalizing session-based recommendations with
  hierarchical recurrent neural networks}. In
  \bibinfo{booktitle}{\emph{Proceedings of the Eleventh ACM Conference on
  Recommender Systems}}. \bibinfo{pages}{130--137}.
\newblock


\bibitem[\protect\citeauthoryear{Rendle, Freudenthaler, and
  Schmidt-Thieme}{Rendle et~al\mbox{.}}{2010}]%
        {FPMC}
\bibfield{author}{\bibinfo{person}{Steffen Rendle}, \bibinfo{person}{Christoph
  Freudenthaler}, {and} \bibinfo{person}{Lars Schmidt-Thieme}.}
  \bibinfo{year}{2010}\natexlab{}.
\newblock \showarticletitle{Factorizing personalized markov chains for
  next-basket recommendation}. In \bibinfo{booktitle}{\emph{Proceedings of the
  19th international conference on World wide web}}. \bibinfo{pages}{811--820}.
\newblock


\bibitem[\protect\citeauthoryear{Sun, Liu, Wu, Pei, Lin, Ou, and Jiang}{Sun
  et~al\mbox{.}}{2019}]%
        {BERT4Rec}
\bibfield{author}{\bibinfo{person}{Fei Sun}, \bibinfo{person}{Jun Liu},
  \bibinfo{person}{Jian Wu}, \bibinfo{person}{Changhua Pei},
  \bibinfo{person}{Xiao Lin}, \bibinfo{person}{Wenwu Ou}, {and}
  \bibinfo{person}{Peng Jiang}.} \bibinfo{year}{2019}\natexlab{}.
\newblock \showarticletitle{BERT4Rec: Sequential recommendation with
  bidirectional encoder representations from transformer}. In
  \bibinfo{booktitle}{\emph{Proceedings of the 28th ACM international
  conference on information and knowledge management}}.
  \bibinfo{pages}{1441--1450}.
\newblock


\bibitem[\protect\citeauthoryear{Tan, Zhang, Yao, Liu, Zhou, Yang, and Hu}{Tan
  et~al\mbox{.}}{2021}]%
        {SINE}
\bibfield{author}{\bibinfo{person}{Qiaoyu Tan}, \bibinfo{person}{Jianwei
  Zhang}, \bibinfo{person}{Jiangchao Yao}, \bibinfo{person}{Ninghao Liu},
  \bibinfo{person}{Jingren Zhou}, \bibinfo{person}{Hongxia Yang}, {and}
  \bibinfo{person}{Xia Hu}.} \bibinfo{year}{2021}\natexlab{}.
\newblock \showarticletitle{Sparse-interest network for sequential
  recommendation}. In \bibinfo{booktitle}{\emph{Proceedings of the 14th ACM
  International Conference on Web Search and Data Mining}}.
  \bibinfo{pages}{598--606}.
\newblock


\bibitem[\protect\citeauthoryear{Tan, Xu, and Liu}{Tan et~al\mbox{.}}{2016}]%
        {GRU4Rec+}
\bibfield{author}{\bibinfo{person}{Yong~Kiam Tan}, \bibinfo{person}{Xinxing
  Xu}, {and} \bibinfo{person}{Yong Liu}.} \bibinfo{year}{2016}\natexlab{}.
\newblock \showarticletitle{Improved recurrent neural networks for
  session-based recommendations}. In \bibinfo{booktitle}{\emph{Proceedings of
  the 1st workshop on deep learning for recommender systems}}.
  \bibinfo{pages}{17--22}.
\newblock


\bibitem[\protect\citeauthoryear{Tang and Wang}{Tang and Wang}{2018}]%
        {Caser}
\bibfield{author}{\bibinfo{person}{Jiaxi Tang} {and} \bibinfo{person}{Ke
  Wang}.} \bibinfo{year}{2018}\natexlab{}.
\newblock \showarticletitle{Personalized top-n sequential recommendation via
  convolutional sequence embedding}. In \bibinfo{booktitle}{\emph{Proceedings
  of the Eleventh ACM International Conference on Web Search and Data Mining}}.
  \bibinfo{pages}{565--573}.
\newblock


\bibitem[\protect\citeauthoryear{Tarvainen and Valpola}{Tarvainen and
  Valpola}{2017}]%
        {2017Mean}
\bibfield{author}{\bibinfo{person}{A. Tarvainen} {and} \bibinfo{person}{H.
  Valpola}.} \bibinfo{year}{2017}\natexlab{}.
\newblock \showarticletitle{Mean teachers are better role models:
  Weight-averaged consistency targets improve semi-supervised deep learning
  results}.
\newblock  (\bibinfo{year}{2017}).
\newblock


\bibitem[\protect\citeauthoryear{Tuan and Phuong}{Tuan and Phuong}{2017}]%
        {3D_CNN}
\bibfield{author}{\bibinfo{person}{Trinh~Xuan Tuan} {and}
  \bibinfo{person}{Tu~Minh Phuong}.} \bibinfo{year}{2017}\natexlab{}.
\newblock \showarticletitle{3D convolutional networks for session-based
  recommendation with content features}. In
  \bibinfo{booktitle}{\emph{Proceedings of the eleventh ACM conference on
  recommender systems}}. \bibinfo{pages}{138--146}.
\newblock


\bibitem[\protect\citeauthoryear{Vaswani, Shazeer, Parmar, Uszkoreit, Jones,
  Gomez, Kaiser, and Polosukhin}{Vaswani et~al\mbox{.}}{2017}]%
        {atten}
\bibfield{author}{\bibinfo{person}{Ashish Vaswani}, \bibinfo{person}{Noam
  Shazeer}, \bibinfo{person}{Niki Parmar}, \bibinfo{person}{Jakob Uszkoreit},
  \bibinfo{person}{Llion Jones}, \bibinfo{person}{Aidan~N Gomez},
  \bibinfo{person}{{\L}ukasz Kaiser}, {and} \bibinfo{person}{Illia
  Polosukhin}.} \bibinfo{year}{2017}\natexlab{}.
\newblock \showarticletitle{Attention is all you need}. In
  \bibinfo{booktitle}{\emph{Advances in neural information processing
  systems}}. \bibinfo{pages}{5998--6008}.
\newblock


\bibitem[\protect\citeauthoryear{Wang, Zhang, Ma, Liu, and Ma}{Wang
  et~al\mbox{.}}{2020}]%
        {wang2020make}
\bibfield{author}{\bibinfo{person}{Chenyang Wang}, \bibinfo{person}{Min Zhang},
  \bibinfo{person}{Weizhi Ma}, \bibinfo{person}{Yiqun Liu}, {and}
  \bibinfo{person}{Shaoping Ma}.} \bibinfo{year}{2020}\natexlab{}.
\newblock \showarticletitle{Make it a chorus: knowledge-and time-aware item
  modeling for sequential recommendation}. In
  \bibinfo{booktitle}{\emph{Proceedings of the 43rd International ACM SIGIR
  Conference on Research and Development in Information Retrieval}}.
  \bibinfo{pages}{109--118}.
\newblock


\bibitem[\protect\citeauthoryear{Wang, Yuan, Chen, Wu, Yang, Sun, and
  Zhang}{Wang et~al\mbox{.}}{2021a}]%
        {StackRec}
\bibfield{author}{\bibinfo{person}{Jiachun Wang}, \bibinfo{person}{Fajie Yuan},
  \bibinfo{person}{Jian Chen}, \bibinfo{person}{Qingyao Wu},
  \bibinfo{person}{Min Yang}, \bibinfo{person}{Yang Sun}, {and}
  \bibinfo{person}{Guoxiao Zhang}.} \bibinfo{year}{2021}\natexlab{a}.
\newblock \showarticletitle{StackRec: Efficient Training of Very Deep
  Sequential Recommender Models by Iterative Stacking}. In
  \bibinfo{booktitle}{\emph{Proceedings of the 44th International ACM SIGIR
  conference on Research and Development in Information Retrieval}}.
  \bibinfo{pages}{357--366}.
\newblock


\bibitem[\protect\citeauthoryear{Wang, Zhu, and He}{Wang
  et~al\mbox{.}}{2021b}]%
        {wang2021cross}
\bibfield{author}{\bibinfo{person}{Jinpeng Wang}, \bibinfo{person}{Jieming
  Zhu}, {and} \bibinfo{person}{Xiuqiang He}.} \bibinfo{year}{2021}\natexlab{b}.
\newblock \showarticletitle{Cross-Batch Negative Sampling for Training
  Two-Tower Recommenders}. In \bibinfo{booktitle}{\emph{Proceedings of the 44th
  International ACM SIGIR Conference on Research and Development in Information
  Retrieval}}. \bibinfo{pages}{1632--1636}.
\newblock


\bibitem[\protect\citeauthoryear{Wu, Li, Hsieh, and Sharpnack}{Wu
  et~al\mbox{.}}{2020}]%
        {wu2020sse}
\bibfield{author}{\bibinfo{person}{Liwei Wu}, \bibinfo{person}{Shuqing Li},
  \bibinfo{person}{Cho-Jui Hsieh}, {and} \bibinfo{person}{James Sharpnack}.}
  \bibinfo{year}{2020}\natexlab{}.
\newblock \showarticletitle{SSE-PT: Sequential recommendation via personalized
  transformer}. In \bibinfo{booktitle}{\emph{Fourteenth ACM Conference on
  Recommender Systems}}. \bibinfo{pages}{328--337}.
\newblock


\bibitem[\protect\citeauthoryear{Wu, Petroni, Josifoski, Riedel, and
  Zettlemoyer}{Wu et~al\mbox{.}}{2019a}]%
        {hard-3}
\bibfield{author}{\bibinfo{person}{Ledell Wu}, \bibinfo{person}{Fabio Petroni},
  \bibinfo{person}{Martin Josifoski}, \bibinfo{person}{Sebastian Riedel}, {and}
  \bibinfo{person}{Luke Zettlemoyer}.} \bibinfo{year}{2019}\natexlab{a}.
\newblock \showarticletitle{Scalable zero-shot entity linking with dense entity
  retrieval}.
\newblock \bibinfo{journal}{\emph{arXiv preprint arXiv:1911.03814}}
  (\bibinfo{year}{2019}).
\newblock


\bibitem[\protect\citeauthoryear{Wu, Tang, Zhu, Wang, Xie, and Tan}{Wu
  et~al\mbox{.}}{2019b}]%
        {SR-GNN}
\bibfield{author}{\bibinfo{person}{Shu Wu}, \bibinfo{person}{Yuyuan Tang},
  \bibinfo{person}{Yanqiao Zhu}, \bibinfo{person}{Liang Wang},
  \bibinfo{person}{Xing Xie}, {and} \bibinfo{person}{Tieniu Tan}.}
  \bibinfo{year}{2019}\natexlab{b}.
\newblock \showarticletitle{Session-based recommendation with graph neural
  networks}. In \bibinfo{booktitle}{\emph{Proceedings of the AAAI Conference on
  Artificial Intelligence}}, Vol.~\bibinfo{volume}{33}.
  \bibinfo{pages}{346--353}.
\newblock


\bibitem[\protect\citeauthoryear{Xiong, Dai, Callan, Liu, and Power}{Xiong
  et~al\mbox{.}}{2017}]%
        {hard-4}
\bibfield{author}{\bibinfo{person}{Chenyan Xiong}, \bibinfo{person}{Zhuyun
  Dai}, \bibinfo{person}{Jamie Callan}, \bibinfo{person}{Zhiyuan Liu}, {and}
  \bibinfo{person}{Russell Power}.} \bibinfo{year}{2017}\natexlab{}.
\newblock \showarticletitle{End-to-end neural ad-hoc ranking with kernel
  pooling}. In \bibinfo{booktitle}{\emph{Proceedings of the 40th International
  ACM SIGIR conference on research and development in information retrieval}}.
  \bibinfo{pages}{55--64}.
\newblock


\bibitem[\protect\citeauthoryear{Xiong, Xiong, Li, Tang, Liu, Bennett, Ahmed,
  and Overwijk}{Xiong et~al\mbox{.}}{2020}]%
        {xiong2020approximate}
\bibfield{author}{\bibinfo{person}{Lee Xiong}, \bibinfo{person}{Chenyan Xiong},
  \bibinfo{person}{Ye Li}, \bibinfo{person}{Kwok-Fung Tang},
  \bibinfo{person}{Jialin Liu}, \bibinfo{person}{Paul Bennett},
  \bibinfo{person}{Junaid Ahmed}, {and} \bibinfo{person}{Arnold Overwijk}.}
  \bibinfo{year}{2020}\natexlab{}.
\newblock \showarticletitle{Approximate nearest neighbor negative contrastive
  learning for dense text retrieval}.
\newblock \bibinfo{journal}{\emph{arXiv preprint arXiv:2007.00808}}
  (\bibinfo{year}{2020}).
\newblock


\bibitem[\protect\citeauthoryear{Xu, Zhao, Liu, Sheng, Xu, Zhuang, Fang, and
  Zhou}{Xu et~al\mbox{.}}{2019a}]%
        {GC-SAN}
\bibfield{author}{\bibinfo{person}{Chengfeng Xu}, \bibinfo{person}{Pengpeng
  Zhao}, \bibinfo{person}{Yanchi Liu}, \bibinfo{person}{Victor~S Sheng},
  \bibinfo{person}{Jiajie Xu}, \bibinfo{person}{Fuzhen Zhuang},
  \bibinfo{person}{Junhua Fang}, {and} \bibinfo{person}{Xiaofang Zhou}.}
  \bibinfo{year}{2019}\natexlab{a}.
\newblock \showarticletitle{Graph Contextualized Self-Attention Network for
  Session-based Recommendation.}. In \bibinfo{booktitle}{\emph{IJCAI}},
  Vol.~\bibinfo{volume}{19}. \bibinfo{pages}{3940--3946}.
\newblock


\bibitem[\protect\citeauthoryear{Xu, Zhao, Liu, Xu, S.~Sheng, Cui, Zhou, and
  Xiong}{Xu et~al\mbox{.}}{2019b}]%
        {RCNN}
\bibfield{author}{\bibinfo{person}{Chengfeng Xu}, \bibinfo{person}{Pengpeng
  Zhao}, \bibinfo{person}{Yanchi Liu}, \bibinfo{person}{Jiajie Xu},
  \bibinfo{person}{Victor S~Sheng S.~Sheng}, \bibinfo{person}{Zhiming Cui},
  \bibinfo{person}{Xiaofang Zhou}, {and} \bibinfo{person}{Hui Xiong}.}
  \bibinfo{year}{2019}\natexlab{b}.
\newblock \showarticletitle{Recurrent convolutional neural network for
  sequential recommendation}. In \bibinfo{booktitle}{\emph{The world wide web
  conference}}. \bibinfo{pages}{3398--3404}.
\newblock


\bibitem[\protect\citeauthoryear{Xu~Xie, Liu, Wu, Gao, Ding, and Cui}{Xu~Xie
  et~al\mbox{.}}{2021}]%
        {CL4SRec}
\bibfield{author}{\bibinfo{person}{Fei~Sun Xu~Xie}, \bibinfo{person}{Zhaoyang
  Liu}, \bibinfo{person}{Shiwen Wu}, \bibinfo{person}{Jinyang Gao},
  \bibinfo{person}{Bolin Ding}, {and} \bibinfo{person}{Bin Cui}.}
  \bibinfo{year}{2021}\natexlab{}.
\newblock \showarticletitle{Contrastive Learning for Sequential
  Recommendation}.
\newblock  (\bibinfo{year}{2021}).
\newblock


\bibitem[\protect\citeauthoryear{Yang, Yi, Zhiyuan~Cheng, Hong, Li,
  Xiaoming~Wang, Xu, and Chi}{Yang et~al\mbox{.}}{2020}]%
        {yang2020mixed}
\bibfield{author}{\bibinfo{person}{Ji Yang}, \bibinfo{person}{Xinyang Yi},
  \bibinfo{person}{Derek Zhiyuan~Cheng}, \bibinfo{person}{Lichan Hong},
  \bibinfo{person}{Yang Li}, \bibinfo{person}{Simon Xiaoming~Wang},
  \bibinfo{person}{Taibai Xu}, {and} \bibinfo{person}{Ed~H Chi}.}
  \bibinfo{year}{2020}\natexlab{}.
\newblock \showarticletitle{Mixed negative sampling for learning two-tower
  neural networks in recommendations}. In \bibinfo{booktitle}{\emph{Companion
  Proceedings of the Web Conference 2020}}. \bibinfo{pages}{441--447}.
\newblock


\bibitem[\protect\citeauthoryear{Yao, Yi, Cheng, Yu, Chen, Menon, Hong, Chi,
  Tjoa, Kang, et~al\mbox{.}}{Yao et~al\mbox{.}}{2020}]%
        {SSL}
\bibfield{author}{\bibinfo{person}{Tiansheng Yao}, \bibinfo{person}{Xinyang
  Yi}, \bibinfo{person}{Derek~Zhiyuan Cheng}, \bibinfo{person}{Felix Yu},
  \bibinfo{person}{Ting Chen}, \bibinfo{person}{Aditya Menon},
  \bibinfo{person}{Lichan Hong}, \bibinfo{person}{Ed~H Chi},
  \bibinfo{person}{Steve Tjoa}, \bibinfo{person}{Jieqi Kang}, {et~al\mbox{.}}}
  \bibinfo{year}{2020}\natexlab{}.
\newblock \showarticletitle{Self-supervised Learning for Large-scale Item
  Recommendations}.
\newblock \bibinfo{journal}{\emph{arXiv preprint arXiv:2007.12865}}
  (\bibinfo{year}{2020}).
\newblock


\bibitem[\protect\citeauthoryear{Yi, Yang, Hong, Cheng, Heldt, Kumthekar, Zhao,
  Wei, and Chi}{Yi et~al\mbox{.}}{2019}]%
        {yi2019sampling}
\bibfield{author}{\bibinfo{person}{Xinyang Yi}, \bibinfo{person}{Ji Yang},
  \bibinfo{person}{Lichan Hong}, \bibinfo{person}{Derek~Zhiyuan Cheng},
  \bibinfo{person}{Lukasz Heldt}, \bibinfo{person}{Aditee Kumthekar},
  \bibinfo{person}{Zhe Zhao}, \bibinfo{person}{Li Wei}, {and}
  \bibinfo{person}{Ed Chi}.} \bibinfo{year}{2019}\natexlab{}.
\newblock \showarticletitle{Sampling-bias-corrected neural modeling for large
  corpus item recommendations}. In \bibinfo{booktitle}{\emph{Proceedings of the
  13th ACM Conference on Recommender Systems}}. \bibinfo{pages}{269--277}.
\newblock


\bibitem[\protect\citeauthoryear{Ying, Zhuang, Zhang, Liu, Xu, Xie, Xiong, and
  Wu}{Ying et~al\mbox{.}}{2018}]%
        {SHAN}
\bibfield{author}{\bibinfo{person}{Haochao Ying}, \bibinfo{person}{Fuzhen
  Zhuang}, \bibinfo{person}{Fuzheng Zhang}, \bibinfo{person}{Yanchi Liu},
  \bibinfo{person}{Guandong Xu}, \bibinfo{person}{Xing Xie},
  \bibinfo{person}{Hui Xiong}, {and} \bibinfo{person}{Jian Wu}.}
  \bibinfo{year}{2018}\natexlab{}.
\newblock \showarticletitle{Sequential recommender system based on hierarchical
  attention network}. In \bibinfo{booktitle}{\emph{IJCAI International Joint
  Conference on Artificial Intelligence}}.
\newblock


\bibitem[\protect\citeauthoryear{Yu, Zhang, Liang, and Zhang}{Yu
  et~al\mbox{.}}{2019}]%
        {MARank}
\bibfield{author}{\bibinfo{person}{Lu Yu}, \bibinfo{person}{Chuxu Zhang},
  \bibinfo{person}{Shangsong Liang}, {and} \bibinfo{person}{Xiangliang Zhang}.}
  \bibinfo{year}{2019}\natexlab{}.
\newblock \showarticletitle{Multi-order attentive ranking model for sequential
  recommendation}. In \bibinfo{booktitle}{\emph{Proceedings of the AAAI
  Conference on Artificial Intelligence}}, Vol.~\bibinfo{volume}{33}.
  \bibinfo{pages}{5709--5716}.
\newblock


\bibitem[\protect\citeauthoryear{Yuan, Karatzoglou, Arapakis, Jose, and
  He}{Yuan et~al\mbox{.}}{2019}]%
        {NextItNet}
\bibfield{author}{\bibinfo{person}{Fajie Yuan}, \bibinfo{person}{Alexandros
  Karatzoglou}, \bibinfo{person}{Ioannis Arapakis}, \bibinfo{person}{Joemon~M
  Jose}, {and} \bibinfo{person}{Xiangnan He}.} \bibinfo{year}{2019}\natexlab{}.
\newblock \showarticletitle{A simple convolutional generative network for next
  item recommendation}. In \bibinfo{booktitle}{\emph{Proceedings of the Twelfth
  ACM International Conference on Web Search and Data Mining}}.
  \bibinfo{pages}{582--590}.
\newblock


\bibitem[\protect\citeauthoryear{Yuan, Duan, Tong, Shi, and Zhang}{Yuan
  et~al\mbox{.}}{2021}]%
        {ICAI-SR}
\bibfield{author}{\bibinfo{person}{Xu Yuan}, \bibinfo{person}{Dongsheng Duan},
  \bibinfo{person}{Lingling Tong}, \bibinfo{person}{Lei Shi}, {and}
  \bibinfo{person}{Cheng Zhang}.} \bibinfo{year}{2021}\natexlab{}.
\newblock \showarticletitle{ICAI-SR: Item Categorical Attribute Integrated
  Sequential Recommendation}. In \bibinfo{booktitle}{\emph{Proceedings of the
  44th International ACM SIGIR Conference on Research and Development in
  Information Retrieval}}. \bibinfo{pages}{1687--1691}.
\newblock


\bibitem[\protect\citeauthoryear{Zhang, Yao, Zhao, Chua, and Wu}{Zhang
  et~al\mbox{.}}{2021b}]%
        {CauseRec}
\bibfield{author}{\bibinfo{person}{Shengyu Zhang}, \bibinfo{person}{Dong Yao},
  \bibinfo{person}{Zhou Zhao}, \bibinfo{person}{Tat-Seng Chua}, {and}
  \bibinfo{person}{Fei Wu}.} \bibinfo{year}{2021}\natexlab{b}.
\newblock \showarticletitle{Causerec: Counterfactual user sequence synthesis
  for sequential recommendation}. In \bibinfo{booktitle}{\emph{Proceedings of
  the 44th International ACM SIGIR Conference on Research and Development in
  Information Retrieval}}. \bibinfo{pages}{367--377}.
\newblock


\bibitem[\protect\citeauthoryear{Zhang, Cheng, Yao, Yi, Hong, and Chi}{Zhang
  et~al\mbox{.}}{2021a}]%
        {MIRec}
\bibfield{author}{\bibinfo{person}{Yin Zhang}, \bibinfo{person}{Derek~Zhiyuan
  Cheng}, \bibinfo{person}{Tiansheng Yao}, \bibinfo{person}{Xinyang Yi},
  \bibinfo{person}{Lichan Hong}, {and} \bibinfo{person}{Ed~H Chi}.}
  \bibinfo{year}{2021}\natexlab{a}.
\newblock \showarticletitle{A Model of Two Tales: Dual Transfer Learning
  Framework for Improved Long-tail Item Recommendation}. In
  \bibinfo{booktitle}{\emph{Proceedings of the Web Conference 2021}}.
  \bibinfo{pages}{2220--2231}.
\newblock


\bibitem[\protect\citeauthoryear{Zhang, Yao, Shao, and Chen}{Zhang
  et~al\mbox{.}}{2019}]%
        {zhang2019nscaching}
\bibfield{author}{\bibinfo{person}{Yongqi Zhang}, \bibinfo{person}{Quanming
  Yao}, \bibinfo{person}{Yingxia Shao}, {and} \bibinfo{person}{Lei Chen}.}
  \bibinfo{year}{2019}\natexlab{}.
\newblock \showarticletitle{NSCaching: simple and efficient negative sampling
  for knowledge graph embedding}. In \bibinfo{booktitle}{\emph{2019 IEEE 35th
  International Conference on Data Engineering (ICDE)}}. IEEE,
  \bibinfo{pages}{614--625}.
\newblock


\bibitem[\protect\citeauthoryear{Zhao, Willemsen, Adomavicius, Harper, and
  Konstan}{Zhao et~al\mbox{.}}{2018}]%
        {zhao2018interpreting}
\bibfield{author}{\bibinfo{person}{Qian Zhao}, \bibinfo{person}{Martijn~C
  Willemsen}, \bibinfo{person}{Gediminas Adomavicius},
  \bibinfo{person}{F~Maxwell Harper}, {and} \bibinfo{person}{Joseph~A
  Konstan}.} \bibinfo{year}{2018}\natexlab{}.
\newblock \showarticletitle{Interpreting user inaction in recommender systems}.
  In \bibinfo{booktitle}{\emph{Proceedings of the 12th ACM Conference on
  Recommender Systems}}. \bibinfo{pages}{40--48}.
\newblock


\bibitem[\protect\citeauthoryear{Zhou, Bai, Song, Liu, Zhao, Chen, and
  Gao}{Zhou et~al\mbox{.}}{2018}]%
        {ATRank}
\bibfield{author}{\bibinfo{person}{Chang Zhou}, \bibinfo{person}{Jinze Bai},
  \bibinfo{person}{Junshuai Song}, \bibinfo{person}{Xiaofei Liu},
  \bibinfo{person}{Zhengchao Zhao}, \bibinfo{person}{Xiusi Chen}, {and}
  \bibinfo{person}{Jun Gao}.} \bibinfo{year}{2018}\natexlab{}.
\newblock \showarticletitle{Atrank: An attention-based user behavior modeling
  framework for recommendation}. In \bibinfo{booktitle}{\emph{Proceedings of
  the AAAI Conference on Artificial Intelligence}}, Vol.~\bibinfo{volume}{32}.
\newblock


\bibitem[\protect\citeauthoryear{Zhou, Liu, Liu, Liu, and Gao}{Zhou
  et~al\mbox{.}}{2017}]%
        {APP-GE}
\bibfield{author}{\bibinfo{person}{Chang Zhou}, \bibinfo{person}{Yuqiong Liu},
  \bibinfo{person}{Xiaofei Liu}, \bibinfo{person}{Zhongyi Liu}, {and}
  \bibinfo{person}{Jun Gao}.} \bibinfo{year}{2017}\natexlab{}.
\newblock \showarticletitle{Scalable graph embedding for asymmetric proximity}.
  In \bibinfo{booktitle}{\emph{Proceedings of the AAAI Conference on Artificial
  Intelligence}}, Vol.~\bibinfo{volume}{31}.
\newblock


\bibitem[\protect\citeauthoryear{Zhou, Ma, Zhang, Zhou, and Yang}{Zhou
  et~al\mbox{.}}{2021}]%
        {CLRec}
\bibfield{author}{\bibinfo{person}{Chang Zhou}, \bibinfo{person}{Jianxin Ma},
  \bibinfo{person}{Jianwei Zhang}, \bibinfo{person}{Jingren Zhou}, {and}
  \bibinfo{person}{Hongxia Yang}.} \bibinfo{year}{2021}\natexlab{}.
\newblock \showarticletitle{Contrastive learning for debiased candidate
  generation in large-scale recommender systems}. In
  \bibinfo{booktitle}{\emph{Proceedings of the 27th ACM SIGKDD Conference on
  Knowledge Discovery \& Data Mining}}. \bibinfo{pages}{3985--3995}.
\newblock


\bibitem[\protect\citeauthoryear{Zhou, Wang, Zhao, Zhu, Wang, Zhang, Wang, and
  Wen}{Zhou et~al\mbox{.}}{2020}]%
        {s3}
\bibfield{author}{\bibinfo{person}{Kun Zhou}, \bibinfo{person}{Hui Wang},
  \bibinfo{person}{Wayne~Xin Zhao}, \bibinfo{person}{Yutao Zhu},
  \bibinfo{person}{Sirui Wang}, \bibinfo{person}{Fuzheng Zhang},
  \bibinfo{person}{Zhongyuan Wang}, {and} \bibinfo{person}{Ji-Rong Wen}.}
  \bibinfo{year}{2020}\natexlab{}.
\newblock \showarticletitle{S3-rec: Self-supervised learning for sequential
  recommendation with mutual information maximization}. In
  \bibinfo{booktitle}{\emph{Proceedings of the 29th ACM International
  Conference on Information \& Knowledge Management}}.
  \bibinfo{pages}{1893--1902}.
\newblock


\end{thebibliography}

\clearpage
\appendix







\end{document}